\documentstyle[preprint,aps,epsfig]{revtex}

\begin{document}
\draft
\tighten

\title{Excited States of Ladder-type Poly-p-phenylene Oligomers}

\author{J\"{o}rg Rissler and Heinz B{\"a}ssler}
\address{Institut f\"{u}r Physikalische Chemie, Kern--Chemie und
Makromolekulare Chemie, Philipps Universit\"{a}t Marburg,
Hans--Meerwein--Stra{\ss}e, D-35032 Marburg, Germany}

\author{Florian Gebhard}
\address{Fachbereich Physik, Philipps Universit\"{a}t Marburg, D-35032 
Marburg, Germany}

\author{Peter Schwerdtfeger}
\address{The University of Auckland, Department of Chemistry, 
Private Bag 92019, Auckland, New Zealand}

\date{{\bf Preprint version of November 23, 2000}}


\maketitle

\begin{abstract}%
Ground state properties and excited states of 
ladder-type paraphenylene oligomers are 
calculated applying semiempirical methods for up to eleven phenylene rings.
The results are in qualitative agreement with experimental data.
A new scheme to interpret the excited states 
is developed which reveals the excitonic 
nature of the excited states. 
The electron-hole pair of the $S_1$-state has a mean distance of approximately 4~{\AA}.
\end{abstract}

\newpage

\section{Introduction}

The investigation of $\pi$-conjugated polymers is in many ways affected by
the structural disorder in these systems.
In contrast, the ladder-type poly-p-phenylenes (LPPP)~\cite{scherf}
offer the opportunity to
study large, rod-like chains of planarised phenylene units.  
As a consequence, the  $\pi$-system might spread out over an entire 
polymer and a vibronic resolution of the 
$\rm S_1\leftarrow S_0$ transition is discernible~\cite{harrison}.
In order to deduce some characteristics of the polymeric films~\cite{olig}, 
like the 
effective conjugation length, several oligomers have been 
synthesized in the past 
to study the low-lying electronic excited states 
of the polymer~\cite{grimme,pauck}.

Photoconduction in LPPP films~\cite{siggi} has been measured as a function
of the energy of the exciting light, too.
A typical small plateau of the 
photocurrent occurs between the absorption 
edge and its steep rise at higher energies and 
extends in this case over 1.6~eV. 
This behavior of the photocurrent 
which does not follow directly the 
absorption features is sometimes called ``abatic''. 

One possible explanation for this effect rests on the 
interpretation of the electronic excited states for the individual 
molecules. Excited states of $\pi$-conjugated 
molecules are usually described as 
Coulomb bound electron-hole pairs. 
This physical picture originates 
from solid-state physics of (organic) semi-conductors. 
Therefore, these molecular states are 
often referred to as excitons, 
although they have to be clearly distinguished from the extended band states 
in a crystal.

A reasonable estimate of the exciton 
binding energy in conjugated 
polymers has been determined, e.g., 
by scanning tunneling spectroscopy measurements~\cite{alvarado} 
which have lead to a value of about $3.5$~eV. 
Excited states with a smaller value, 
and larger electron-hole distance, respectively, 
should be more susceptible to the 
separation via an external electric field. 
Following this idea, the conjecture has been brought forward
that such a state is responsible 
for the steep rise of the photocurrent in 
poly-phenylene-vinylene (PPV)~\cite{bredas}. 
Later on, another explanation has followed based 
on the excess photon energy 
which is converted to the vibrational heat bath~\cite{archipov}. 
The latter proposal is now widely accepted. 

In order to test these concepts for excited states of $\pi$-conjugated 
systems, several oligomers of the LPPP type with up to 
eleven phenylene rings are investigated in this work. 
The study of oligomers instead of an (infinite) polymer 
follows the above mentioned approach and allows the direct 
comparison with experiment. The main difference to the 
experiments in condensed phases is the 
restriction to single chains in the vacuum.

As the experimentally used molecules are computationally too demanding
one has to replace the large
aliphatic substituents attached to LPPP 
by hydrogen (see Figure~\ref{Fig:1} and Table~\ref{Tab:0}). 
This should have only negligible 
effects on the optical properties, which are 
governed by the frontier orbitals of $\pi$-symmetry.
These aliphatic substituents are only necessary for the better
solubility of the polymer, or to prohibit the formation of 
aggregates in the film ($\rm R_3$ = methyl in Figure~\ref{Fig:1}).
 
%
%

Since the systems studied here reach the size of the 
effective conjugation length proposed 
for LPPP (about 14 phenylene rings~\cite{grimme,harrison}), 
ab-initio or density functional methods are not applicable, and 
one has to assent to less expensive 
semiempirical methods (AM1, INDO/S; see below). 
Thus, the wave functions of the corresponding ground states
are the INDO (intermediate neglect of 
differential overlap) Slater determinants
$|\Phi_0^{\rm INDO}\rangle$~\cite{pople}.
For the excited states $\left|\Phi_s^{\rm CIS}\right\rangle $ (see equation \ref{PHICISdef}), the INDO/S expansion is used in the 
spin-singlet sector.
The excited states with dominant oscillator strengths
will be addressed as $\rm S_1$ 
for the first excited state, $\rm S_m$ 
for the intermediate excited state and $\rm S_F$ for the 
high energy, ``Frenkel-type'' state. 
The electronic ground state will be denoted as $\rm S_0\equiv 
|\Phi_0^{\rm INDO}\rangle$. 

The article is organized as follows. In Sect.~\ref{Methods},
the semiempirical approach is briefly described.
In Sect.~\ref{GeometrySpectra}, the results for the 
geometric structure of the oligomers and their spectra are presented.
The main part of this article, Sect.~\ref{Interpretation}, focuses on the 
development of a general interpretation scheme 
for excited-state wave functions. Its application to 
INDO/S wave functions leads in a 
straightforward way to the interpretation of the excited states as 
bound electron-hole pairs. A short conclusion closes the presentation.

\section{Methods}
\label{Methods}

Although it is not feasible to calculate the higher oligomers by
first-principle methods, the oligomer with two phenylene rings ($n=0.5$) has
been calculated at MP2/6-31G* level~\cite{MP2,basis}(M{\o}ller-Plesset Pertubation Theory of second order). 
The results are used as a reference for the semiempirical methods. 

Following a procedure of Br\'edas {\it et al.}~\cite{bredas2}, 
one has to choose a semiempirical method 
which yields reliable geometric structures. 
In the present case the Austin semiempirical parametrization (AM1)~\cite{dewar} 
gives an almost exact agreement with the MP2 results (cf.~$\sum|\Delta r_{\rm C-C}|$ 
for the bond lengths in Table~\ref{Tab:1}).
This method will therefore be used to calculate 
the geometric structure of the ground states.
Note, however, that the 
PM3 method~\cite{stewart} yields better results for zero-point vibrational energies
(ZPE). 

The AM1 and MP2 calculations have been carried out on a IBM/SP2
computer using the GAUSSIAN94 (Rev.~D4)~\cite{gauss} program package. 
All minima are characterized by diagonalization of the
Hessian Matrix, whereby the zero-point energies (ZPE) have been obtained,
too. The latter will be given unscaled throughout the paper.

In the next step, excited singlet states are calculated 
using Zerner's INDO/S method~\cite{zerner} based on the 
minimum AM1 structures from the electronic ground states.
Thus it is clear that geometric relaxation effects
in the excited state are neglected.
The active CI space consists of the 22 highest occupied and the 22 lowest
unoccupied molecular orbitals. This is the biggest active space
possible within the used program package and it contains orbitals
of $\sigma$ symmetry for some oligomers as well. 
As expected, the dominant configuration state functions
in the wave function describe $\pi$--$\pi^*$ excitations.
The calculation of the spectra have been accomplished on a
PentiumIII--PC using the program package CAChe~3.1~\cite{cache}.

In order to get a more realistic view of 
the calculated line spectra, 
Gaussian peaks are least-square fitted 
to the INDO/S oscillator strengths.
This procedure masks transitions with moderate oscillator strengths 
which are close in energy to a dominant transition. 
This is especially the case for the high-energy 
region of the spectrum, in which strong 
$\rm S_F\leftarrow S_0$ transitions cover others. 
The hidden states are not visible in the optical spectrum, and
consist of a number of important configurations sensitive to the
size of the active CI space (Figure~\ref{Fig:2}).
The remaining part of the spectra comprises only features with one transition, 
except in two cases, where two almost identical 
states are close together. In these cases 
only one state will be discussed (the $\rm S_1\leftarrow S_0$ transition 
for the $n=0.5$ oligomer and the $\rm S_m\leftarrow S_0$ transition 
for the trimer; see discussion below).

\section{Results for geometry and optical spectra}
\label{GeometrySpectra}

\subsection{Geometry}

All structures have been optimized without any symmetry constraints.
Therefore, the geometries only fit to certain
point groups within crystallographic accuracy ($\pm$ 0.001~{\AA}, $\pm
2^{\circ}$). The resulting point groups are equivalent to ones 
suggested by the simple Lewis-type structure, see Figure~\ref{Fig:1}. 
Oligomers with an even number of phenylene rings adopt $C_{2v}$ symmetry. 
Those with an odd number adopt $C_{2h}$ symmetry. The C--C bond lengths 
range between about 1.380~{\AA}
and about 1.500~{\AA}. Figure~\ref{Fig:1.5} shows the MP2 result for the 
$n = 0.5$ oligomer as an example.

As the smallest C--C bond distance in a 
methylene bridge is 1.503~{\AA} long, they are assigned to single bonds.
As a consequence, the hydrogen atoms
of the methylene bridge do not participate
in the conjugated system, although this would be allowed. This supports the validity of neglecting
the aliphatic substituents.


\subsection{Optical spectra}

In Figure~\ref{Fig:2}
the  (vibronically unresolved)
calculated spectra of LPPP oligomers are plotted.
A comparison with measured fluorescence spectra of different 
oligomers~\cite{pauck} shows qualitative agreement, i.e.,
they show a broad $\rm S_1\leftarrow S_0$ transition which is shifting to
lower energies with increasing system size.
As one would expect for the $\rm S_1$ state, the HOMO--LUMO 
excitation is the dominant one. HOMO (LUMO) 
refers to the highest occupied (lowest unoccupied)
molecular orbital in the INDO ground state.

A second interesting feature
can be seen in the calculations for
the oligomers with five and more rings. It will be referred to as the $\rm
S_m\leftarrow S_0$ transition, because it is 
energetically in the middle between the $\rm S_1$ state and the high 
energy region. In the $\rm S_m$
state the dominant determinants are built by the substitution of
the HOMO by the LUMO+2 and the substitution of the HOMO$-$1 by the
LUMO+1. The position of the $\rm S_m\leftarrow S_0$ transition also shifts
to lower energies with increasing oligomer length.
This is not discernible in the experimental
spectra because of the small corresponding oscillator strengths.

Finally, both
sets of spectra show a steep rise at around 5.40~eV which is dominated by
a transition at approximately 5.85~eV for every oligomer. 
This transition will be called 
the $\rm S_F\leftarrow S_0$ transition,
as the $\rm S_F$ state shows a high degree of localization 
(i.e., a ``Frenkel'' state), which will become clear later in the discussion.
The $\rm S_F$ state is determined by several configuration state
functions, where low-lying occupied molecular orbitals are
exchanged by orbitals with high energy.

In Figure~\ref{Fig:3a} the energy of all three 
optically important transitions is drawn as a
function of the reciprocal number of phenylene rings in the molecule.
The transitions of the $\rm S_1$ and the $\rm S_m$ states
show a strong linear dependence, 
whereas the energy of the $\rm S_F\leftarrow S_0$ transitions hardly changes
with system size, in agreement with the experimental observation.
Despite this qualitative agreement, 
a quantitative comparison with the experimental 
values for the effective conjugation length~\cite{grimme} is not possible, 
since the theoretical curve leads to an unphysical negative value.
The main reason for this error is the 
neglect of the polarization energy in the calculations
which stabilizes the excited states in the condensed phases in the experiments.
Moreover, the linear extrapolations for the transitions to the 
$\rm S_m$ and the $\rm S_1$ states 
suggest a crossing of the two energies
which must not be taken for granted at this stage.


{}From the electronic dipole transition moments $\mathbf{M}$ for 
the $\rm S_1\leftarrow S_0$ transitions, 
one obtains an estimate for the radiative 
fluorescence lifetimes $\tau$ for every oligomer, using
\begin{mathletters}
\label{lifetime}
\begin{eqnarray}
\mathbf{M} &=& e 
\left\langle \Phi_s^{\rm CIS} \left| 
\sum_i z_i \vec{r}_i\right| \Phi_0^{\rm INDO}\right\rangle \; , \\ 
\tau &=& \frac{3hc^3}{8\pi\omega^3|\mathbf{M} |^2} \; .
\end{eqnarray}\end{mathletters}%
Here, $e$ is the elementary charge, $\vec{r}_i$ is 
the position vector of the {\it i}-th particle of charge $z_i e$,
$h$ is Planck's constant, $c$ is the speed of light, and
$\hbar\omega$ is the energy of the transition, all
quantities measured in cgs units.
The lifetimes $\tau$ show a linear 
dependence on the reciprocal number of 
phenylene rings, see Figure~\ref{Fig:3b}. Here, the value of 
several hundred picoseconds for the large 
oligomers is in quantitative 
agreement with the experimental value of about 300~ps for the polymer, 
measured in an organic matrix~\cite{iura}.
Under the assumption of a linear relation between~$\tau$ and the 
inverse of the oligomer length, one finds from the data in Table~\ref{Tab:3} that 
a value of $\tau_{\rm polymer}=300\, {\rm ps}$ 
corresponds to an effective conjugation length of
about 20~phenylene rings, in reasonable agreement with
the value of 14~rings estimated from other optical 
experiments~\cite{grimme,harrison}. 
 
\section{Interpretation}
\label{Interpretation}
\subsection{General considerations}

For an interpretation the wave functions for the 
ground state and the excited states 
need to be related to experimentally 
observable quantities. 
The optical absorption at frequency $\omega$ 
is proportional to the real part of the optical conductivity, 
as given by the Kubo formula~\cite{mahan}
\begin{eqnarray}
{\rm Re\,} \left[ \sigma(\omega>0) \right]= \frac{{\rm Im\,} \left[ \chi_{jj}(\omega >0)\right] }{\omega}
\; ,
\end{eqnarray}
where $\chi_{jj}(\omega)$ is the current--current correlation function,
\begin{eqnarray}
\chi_{jj}(\omega >0) & = & -\left\langle\Phi_0\left|
\hat{\jmath}\frac{1}{E_0-\hat{H}+\omega+i\eta}\hat{\jmath}
\right|\Phi_0\right\rangle\nonumber\\
  & = & -\sum\limits_{\left|\Phi_s\right\rangle}
\frac{\left|\left\langle\Phi_s\left|\hat{\jmath}
\right|\Phi_0\right\rangle \right|^2}{\omega-(E_s-E_0)+i\eta} 
\; .
\label{chi}
\end{eqnarray} 
Here, $\hat{H}$ is the Hamilton operator of the system, 
$\left|\Phi_s\right\rangle$ are its eigenstates 
with energies $E_s$ ($s = 0, 1, 2,\ldots$), $\hbar\equiv 1$,
and $\eta=0^+$ is positive infinitesimal. Therefore, the real part of 
the optical conductivity may be written as
\begin{equation}
{\rm Re\,} \left[\sigma(\omega>0) \right]= 
\frac{\pi}{\omega}\sum\limits_{\left|\Phi_s\right\rangle}
\left|\left\langle\Phi_s\left|\hat{\jmath}
\right|\Phi_0\right\rangle\right|^2 \;\delta\left(\omega-(E_s-E_0)\right) \; .
\label{Resigmadelta}
\end{equation}
The current operator is defined by
\begin{eqnarray}
\hat{\jmath} = \sum_{\sigma}
\sum_{m,n} p_{m,n} \hat{C}^{\dagger}_{n,\sigma} \hat{C}_{m,\sigma}\; ,
\end{eqnarray}
where $\hat{C}^{\dagger}_{n,\sigma}$, $\hat{C}_{n,\sigma}$ are creation and 
annihilation operators for 
electrons with spin~$\sigma=\uparrow,\downarrow$ 
in the molecular orbitals $\phi_n$, and $p_{m,n}$ is the 
matrix element between the corresponding one-particle states.

Equation~(\ref{Resigmadelta}) is readily interpreted. 
The absorption of a photon with energy $\omega$ 
induces an excitation between the ground state $\left|\Phi_0\right\rangle$ 
and the excited states  $\left|\Phi_s\right\rangle$ with 
energy $\omega = E_s-E_0$. The amplitude for this absorption process
\begin{eqnarray}
A_{0,s} = \left\langle\Phi_s\left|\hat{\jmath}\right|\Phi_0\right\rangle
\end{eqnarray}
determines the oscillator strength, 
$f_{0,s} \propto \left| A_{0,s} \right|^2$.

For a further analysis of the amplitudes $A_{0,s}$,
the current operator is expressed in terms of the field operators,
\begin{mathletters}
\begin{eqnarray}
\hat{\Psi}_{\sigma}^{\dagger}(\vec{x}) &=& 
\sum\limits_n \phi^*_n(\vec{x}) \hat{C}^{\dagger}_{n,\sigma}\; , \\
\hat{\Psi}_{\sigma}(\vec{x}) &=& \sum\limits_n \phi_n(\vec{x}) 
\hat{C}_{n,\sigma} \; ,
\end{eqnarray}\end{mathletters}%
which create/annihilate an electron with spin~$\sigma$ at $\vec{x}$. From 
the completeness of the molecular orbitals $\phi_n(\vec{x})$ 
one readily obtains
\begin{eqnarray}
\hat{\jmath} = \sum_{\sigma} \int d\vec{x} d\vec{y} \; j(\vec{x}, \vec{y})  
\hat{\Psi}_{\sigma}^{\dagger}(\vec{x})  \hat{\Psi}_{\sigma}(\vec{y}) 
\end{eqnarray}
with
\begin{eqnarray}
j( \vec{x} , \vec{y}) = \sum\limits_{n,m} 
p_{m,n}\phi_n(\vec{x}) \phi^*_m(\vec{y})\; . 
\label{matrixelement}
\end{eqnarray}
Therefore, the amplitudes can be cast into the form
\begin{eqnarray}
A_{0,s} = \sum_{\sigma} \int d\vec{x}_e d\vec{x}_h \; 
j(\vec{x}_e, \vec{x}_h) A^{\sigma}_{0,s}(\vec{x}_e, \vec{x}_h) \; , 
\end{eqnarray}
where the particle-hole amplitudes 
\begin{eqnarray}
A^{\sigma}_{0,s}(\vec{x}_e, \vec{x}_h ) = \left\langle\Phi_s
\left| \hat{\Psi}_{\sigma}^{\dagger}(\vec{x}_e) 
\hat{\Psi}_{\sigma}(\vec{x}_h)\right|\Phi_0\right\rangle
\label{amplitude}
\end{eqnarray}
are given by the overlap between the excited state 
$\left|\Phi_s \right\rangle$ 
and the ground state $\left|\Phi_0 \right\rangle$ 
with an electron at $\vec{x}_e$ and 
a hole at  $\vec{x}_h$ (if $\vec{x}_e \not=  \vec{x}_h$).

It is thus seen that the quantities $A^{\sigma}_{0,s}(\vec{x}_e, \vec{x}_h )$ 
allow to address the question in how far a 
given excited state  $\left|\Phi_s \right\rangle$ may be viewed 
as an electron-hole excitation of the ground state, 
and they directly enter the Kubo formula 
for the optical conductivity. Note that the analysis of~(\ref{amplitude}) 
does not require the full 
spatial dependence of the many-particle wave functions  
$\left|\Phi_s \right\rangle$ or  $\left|\Phi_0 \right\rangle$  
as in~\cite{bredas,gw} or the density--density 
correlation function of the ground state as in~\cite{mazum}.

The matrix elements for the current operator, 
$j( \vec{x}_e , \vec{x}_h )$ in~(\ref{matrixelement}),
do not change much over atomic distances. 
Therefore, it is usually sufficient to 
introduce the coarse-grained 
densities for the electron-hole content of $\left|\Phi_s\right\rangle$
with respect to~$|\Phi_0\rangle$,
\begin{eqnarray}
P_{0,s}(i, j) = \sum_{\sigma} \int d\vec{x}_e d\vec{x}_h 
\left|  A^{\sigma}_{0,s}(\vec{x}_e, \vec{x}_h )   \right|^2 \; 
\Theta(\vec{x}_h - \vec{r}_i)\Theta(\vec{x}_e - \vec{r}_j)\; ,
\label{Pij}
\end{eqnarray}
where
\begin{eqnarray}
\Theta(\vec{x}-\vec{r}_i)=\left\{
\begin{array}{rcl}
1 & \mbox{,}& \mbox{if $\vec{x}-\vec{r}_i \in V_i$} \\ 
0 & \mbox{,} & \mbox{else} \\ \end{array}
\right.
\end{eqnarray}
is the step function for the (atomic) volume $V_i$ 
around atom~$i$. $P_{0,s}(i, j)$ is the overlap density
between the excited~$|\Phi_s\rangle$ and the ground state~$|\Phi_0\rangle$
with an electron-hole pair around the nuclei at~$i$ and~$j$, respectively.

Equation~(\ref{amplitude}) is exact and 
applies to all quantum-mechanical systems. 
It is currently under investigation for the 
numerical analysis of quantum lattice systems 
with the density matrix renormalisation group (DMRG) method~\cite{eric}. 
In the next subsection, 
the present case of INDO/S wave functions will be studied in more detail.

\subsection{Application to INDO/S}

As described earlier, the ground state is approximated 
as an INDO Slater determinant
\begin{mathletters}
\begin{eqnarray}
\left|\Phi_0\right\rangle &\approx&  \left|\Phi^{\rm INDO}_0\right\rangle 
= \prod_{\sigma} \left|\Phi^{\rm INDO}_{\sigma, 0}\right\rangle =
\prod_{\sigma} \prod_{1\leq a\leq n} \hat{C}_{a,\sigma}^{\dagger} 
|{\rm vacuum}\rangle \; , \\
\left|\Phi^{\rm INDO}_{\sigma, 0}\right\rangle 
&=& \frac{1}{n!}  \int d\vec{x}_1\ldots d\vec{x}_n
{\rm Det} \left[ {\phi}_a(\vec{x}_i) \right]
\Psi_{\sigma}^{\dagger}(\vec{x}_1) \ldots \Psi_{\sigma}^{\dagger}(\vec{x}_n)
|{\rm vacuum}\rangle \; ,
\end{eqnarray}\end{mathletters}%
where the molecular orbitals $\phi_a(\vec{x}_i)$ ($1\leq a\leq n$) are 
expressed as a linear combination of spatial 
atomic orbitals (AOs) $\chi_b(\vec{x})$ 
\begin{equation}
\phi_a(\vec{x}_i)=\sum\limits_{b}^{\rm AOs} d^{a}(b) \chi_b(\vec{x}_i) \; .
\label{MO}
\end{equation}

The atomic orbitals are centered at certain nuclei such that $b \equiv (\vec{r}_b, \beta_b)$ comprises the orbital type ($\beta_b = s, p, d\ldots$) and its position $\vec{r}_b$.

The spin singlet excited-state wave functions 
are described as linear combinations of 
singly-excited configuration state functions
\begin{mathletters}
\label{PHICISdef}
\begin{eqnarray}
\left|\Phi_s\right\rangle &\approx& \left|\Phi_s^{\rm CIS}\right\rangle\; ,\\
\left|\Phi_s^{\rm CIS}\right\rangle &=&
\sqrt{\frac{1}{2}}\sum_{\sigma}
\sum\limits_{a,r}c_{a,r} \hat{C}_{r,\sigma}^{\dagger}  
\hat{C}_{a,\sigma}   \left|\Phi^{\rm INDO}_0\right\rangle \; , 
\end{eqnarray}\end{mathletters}%
where the indices $a$ and $r$ refer to the active space of occupied and 
virtual molecular orbitals, respectively.

With these approximations for $\left|\Phi_0\right\rangle$ and  
$\left|\Phi_s\right\rangle$, it is obvious that the 
coordinates of the added electron enter~(\ref{amplitude}) 
only through the virtual MOs~$\phi_r$ whereas the hole is 
described by the eliminated, occupied MOs $\phi_a$,
\begin{eqnarray}
A^{\sigma}_{0,s}(\vec{x}_e, \vec{x}_h ) & \approx&   
\left\langle \Phi_s^{\rm CIS}\left|\hat{\Psi}_{\sigma}^{\dagger}(\vec{x}_e)
\hat{\Psi}_{\sigma}(\vec{x}_h) \right|
\Phi_0^{\rm INDO} \right\rangle \nonumber \\
&=& \sqrt{\frac{1}{2}}
\sum\limits_{a,r} c_{a,r}^*\phi^*_r(\vec{x}_e) \phi_a(\vec{x}_h)
\; .
\end{eqnarray}
To make further progress, one has to address the coarse-grained 
densities $P_{0,s}(i, j)$ in~(\ref{Pij}), 
\begin{eqnarray}
P_{0,s}(i, j) & \approx &  P_{0,s}^{\rm INDO/S}(i, j)\nonumber \\
P_{0,s}^{\rm INDO/S}(i, j) & = &\int d\vec{x}_e d\vec{x}_h 
\left| \sum\limits_{a,r} c_{a,r}^* \phi^*_r(\vec{x}_e) \phi_a(\vec{x}_h)
 \right|^2 \; 
\Theta(\vec{x}_h - \vec{r}_i)\Theta(\vec{x}_e - \vec{r}_j) \\
&=& \sum\limits_{\beta_i, \beta_j} \sum\limits_{ {{a, r} \atop  {a', r'}}}  
c_{a,r}^{\ast} c_{a', r'}^{} (d^r(\vec{r}_j ,\beta_j))^\ast d^{r'}(\vec{r}_j ,\beta_j) (d^{a'}(\vec{r}_i ,\beta_i))^{\ast} d^a(\vec{r}_i ,\beta_i)
 \; .
\nonumber
\end{eqnarray}
In the last step the INDO approximation~\cite{pople}
and the orthogonality of the atomic orbitals on the same atom were used, i.e.,
\begin{equation}
\int d\vec{x} \; \Theta(\vec{x}-\vec{r}_j) \chi_{b}^*(\vec{x})
\chi_{b'}^{}(\vec{x}) \approx \delta_{b,b'}\delta_{\vec{r}_b,\vec{r}_j} \; .
\end{equation} 
Recall that $d^r(\vec{r}_j, \beta_j)$ denote AO coefficients, and $c_{a,r}$ 
are the CI coefficients of the CIS wave function.
In this case, one may verify that $\sum_{i,j} P_{s,0}(i,j) = 1$, 
if one takes into account the INDO approximation and the normalization of
$|\Phi_s^{\rm CIS}\rangle$ in~(\ref{PHICISdef}),
\begin{equation}
1= \langle \Phi_s^{\rm CIS}| \Phi_s^{\rm CIS}\rangle
= \int \left| \sum\limits_{a,r} c_{a,r}^*\phi^*_r(\vec{x}_e) \phi_a(\vec{x}_h)
\right|^2\,d\vec{x}_e d\vec{x}_h = 
\sum\limits_{i, j}^{{\rm nuclei}} P^{\rm INDO/S}_{s,0}(i,j)\; .
\end{equation}
Therefore, $P^{\rm INDO/S}_{s,0}(i,j)$ may be viewed as a discrete
probability function for finding a hole around the nucleus~$i$ and an
electron around the nucleus~$j$.
This simplification is only valid in the INDO/S approximation.

A check can be made, whether the values of $P^{\rm INDO/S}_{s,0}(i,j)$
carry significant weight corresponding to AOs of $\sigma$~symmetry.
It turns out that this weight is vanishingly small, and one can sum all
contributions of the AOs of an atom without losing information.

One has to verify that the detected electron-hole correlations for an
excited state really stem from the electron-hole interaction
rather than from a coincidence of the motion of two independent 
particles confined to a small molecule. Therefore,
the quantity
\begin{equation}
A_{0,{\rm H\to L}}^{\sigma}=
\left\langle \Phi_{\rm H \to L} \left|\hat{\Psi}_{\sigma}^{\dagger}(\vec{x}_e)
\hat{\Psi}_{\sigma}(\vec{x}_h)  \right| \Phi_0^{\rm INDO} \right\rangle
\label{HOMOLUMO}
\end{equation}
has been
investigated according to the scheme~(\ref{amplitude}). Here,
$|\Phi_{\rm H \to L}\rangle$ is the INDO wave function where the
HOMO is substituted by the LUMO. This corresponds to the pure
``band-like'' excitation, i.e., in a CIS state representation this
configuration state function has the coefficient $c_{a,r}=1$
for $a={\rm HOMO}$, $r={\rm LUMO}$, and all
other coefficients are set to zero.

Given the $P^{\rm INDO/S}_{s,0}(i,j)$ 
an expectation value 
$\langle E \rangle$ of the distance of the 
electron-hole pair can be calculated with
the help of the bond lengths $r(i,j) = 
|\vec{r}_i - \vec{r}_j| $ of the molecule,
\begin{eqnarray}
\langle E\rangle_{s,0}  = \sum\limits_{i, j} P^{\rm INDO/S}_{s,0}(i,j)\;
r(i,j) \; .
\end{eqnarray}
The standard deviation $\sigma_{s,0}$ is equally accessible,
\begin{eqnarray}
\left(\sigma_{s,0}\right)^2 = \sum\limits_{i, j} \left( r(i,j)
-\langle E\rangle_{s,0} \right)^2 P^{\rm INDO/S}_{s,0}(i,j)
 \; .
\end{eqnarray}
In small molecules, the confinement of the oligomers will 
determine the distance of the electron-hole pair. 
In larger molecules, 
if the interaction between hole and electron
is weak, $\langle E\rangle_{s,0}$ and $\sigma_{s,0}$ 
will increase with system size  
since the particles move essentially independently through the molecule.
On the other hand, 
the interaction between electron and 
hole may keep them together at a fixed distance even though the 
size of the molecule increases. Such a
bound electron-hole pair 
may lead to a constant value of $\langle E\rangle_{s,0}$ and $\sigma_{s,0}$ 
for every system size. Note, however,
$\langle E\rangle_{s,0}$ and $\sigma_{s,0}$
only contain an information about the overall extension of an excitation but
may fail to describe the localization of an
electron-hole pair onto segments of the molecule;
an example of this situation is given below.


In order to get a more pictorial way of the electron-hole pair distribution,
one may concentrate on a quasi one-dimensional 
chain of carbon atoms; Figure~\ref{Fig:4a} shows how those 
chains are chosen. When $P_{s,0}(i,j)$ is
plotted for this chain a bound electron-hole pair will show large values
along the diagonal of the plot and vanishingly small values
in the off-diagonal regions. Unbound pairs will lead to the
opposite situation.

The two-dimensional distribution $P_{s,0}(i,j)$ can be further 
smoothed into
\begin{equation}
\overline{P}_{s,0}(r)= \int_{r-\Delta r/2}^{r+\Delta r/2}
dr' \sum_{i,j}^{\rm nuclei} P_{s,0}(i,j) \: \delta\left(
r'-|\vec{r}_i-\vec{r}_j|\right) \; ,
\label{smoothedP}
\end{equation}
which is solely a function of the electron-hole distance.
The choice of $\Delta r =1.8\, \hbox{\AA}$ gives smooth curves
as a function of~$r$.
$\overline{P}_{s,0}(r)$ gives 
the most concise description of the electron-hole excitation.

\subsection{Results}

The $P_{s,0}^{\rm INDO/S}(i,j)$ matrix
has been calculated for the states,
whose transitions are of significant oscillator strength in the
spectra: The $\rm S_1$, the $\rm S_m$, and the $\rm S_F$ state.

First, the state $\rm S_1$ is discussed.
As seen from Figure~\ref{Fig:4b}, no bound state is formed for only
two phenylene rings. In fact, 
the molecular confinement is 
dominant up to a chain length of
four phenylene rings. 
For five phenylene rings, see Figure~\ref{Fig:4c}, and larger
systems, a bound electron-hole pair is discernible.
The shape of $P_{{\rm S_1},0}(i,j)$ 
in the plot along the one-dimensional chain does
not change for systems larger then five phenylene rings.
The $\rm S_1$ state clearly corresponds to a strongly bound 
electron-hole pair.

Figure~\ref{Fig:5a} shows the average electron-hole separation and 
the corresponding standard deviation for the $\rm S_1$~state
as a function of system size. They saturate for more than five phenylene
rings, $\langle E\rangle_{{\rm S_1},0}\approx 4\, \hbox{\AA}$. 
This is in reasonable agreement with the experimental 
value of 7~{\AA} obtained from electro-absorption 
measurements~\cite{harrison}. 
Some results of electron energy-loss spectroscopy, however, 
seem to hint at a totally different 
behavior of the excited states: they are supposed 
always to increase with increasing length of the oligomer~\cite{knupfer}.
At present, the reason for this discrepancy is not clear.

The saturation behavior for $\langle E\rangle_{{\rm S_1},0}$ and
the apparently large saturation value for the standard 
deviation~$\sigma_{{\rm S_1},0} \approx 3\, \hbox{\AA}$ are readily 
understood from
the ``smoothed'' probability distribution
$\overline{P}_{{\rm S_1},0}(r)$, see~(\ref{smoothedP}).
As seen in Figure~\ref{Fig:5b}, $\overline{P}_{{\rm S_1},0}(r\rightarrow 0)$ 
decreases with increasing system size, the whole distribution broadens,
 and develops a maximum around $r_m=4\, \hbox{\AA}$.
For oligomers with five and more phenylene rings the 
distribution does not change significantly. 
Since at the same time the system grows, 
the values of $\langle E\rangle_{{\rm S_1},0}$ and $\sigma_{{\rm S_1},0}$ 
saturate.
The relatively large values of $\sigma_{{\rm S_1},0}$ are due to
the location of the maximum of the distribution at a finite value of $r_m$.
Finally, Figure~\ref{Fig:5c} shows the probability distribution
$P_{S_1,0}^{\rm INDO/S}(i,j)$ for the largest oligomer with
eleven phenylene rings ($n=5$).

Next, the state $\rm S_m$ is addressed.
Br\'edas and coworkers have  located a state in the 
calculated spectrum of poly-p-phenylene-vinylene (PPV), which they assign to a charge transfer state. 
In this context this state is  
equivalent to an unbound electron-hole pair. 
The optical transition to this charge transfer state is in the same 
region as the steep increase of the photocurrent in the respective 
polymeric film. As a result, they conclude that the population of this 
state is responsible for this ``abatic'' behavior of the photocurrent. 
The $\rm S_m$ state obtained in the present calculation
lies in the same region as the 
charge-transfer state in PPV, that is between the $\rm S_1$ 
and the high-energy states.
The question is, whether the $\rm S_m$ state 
is a charge transfer state or not.
 
As seen in Figure~\ref{Fig:6a}, the $\rm S_m$ state shows 
almost constant values of $\langle E\rangle_{{\rm S_m},0}$ 
and $\sigma_{{\rm S_m},0}$.  
A rise in these values occurs for the $n=5$ oligomer which hints at 
unbound states for larger systems.
This idea, however, is
not supported by a more detailed look at the graphs for 
$\overline{P}_{{\rm S_m},0}(r)$ as a function of oligomer length,
see Figure~\ref{Fig:6b}, and the results for 
$P_{{\rm S_m},0}(i,j)$ for the largest oligomer, see Figure~\ref{Fig:6c}. 
The curves are qualitatively the same as for the corresponding
$\rm S_1$ states. For example, 
compare Figures~\ref{Fig:5c} and~\ref{Fig:6c}: $P_{{\rm S_1},0}(i,j)$ and
$P_{{\rm S_m},0}(i,j)$ are essentially zero 
in the off-diagonal regions. Similarly, 
the curves $\overline{P}_{{\rm S_m},0}(r)$ as a function of oligomer length
resemble those of 
$\overline{P}_{{\rm S_1},0}(r)$, compare Figures~\ref{Fig:5b} and~\ref{Fig:6b}.
The only difference to the
$\rm S_1$ state is the localization of the excitation to three
parts of the molecule in $\rm S_m$.
Because of this, one cannot interpret this
state as an unbound elctron-hole pair or \~ exciton, and consequently no explanation for
the rise of the photocurrent of LPPP at about 4.0~eV can be 
given at this level of
theory. This is in line with the assumption 
that local heating due to excess energy 
is the reason for the behavior of the photocurrent~\cite{archipov} 
and not the nature of the optically accessible states.

Lastly, for the $\rm S_F$ state, 
the mean electron-hole separation does not saturate but grows with system size,
see Fig.~\ref{Fig:7a}. From such an analysis one might conclude
a delocalized state.
On the contrary, for the $\rm S_F$ state the degree of localization is
{\sl higher\/} than in the other two excited states, and 
the electron-hole pair is actually restricted to every single
phenylene ring.
As can be seen from Figures~\ref{Fig:7b} and~\ref{Fig:7c}, the 
apparent lack of convergence in $\langle E\rangle_{{\rm S_F},0}$
and $\sigma_{{\rm S_F},0}$ relates to the fact that isolated elctron-hole pairs
on chain segments can be linearly superimposed, i.e., an extended state can be formed which, nevertheless,
does not contribute to the conductivity.
As for the states $\rm S_1$ and $\rm S_m$, 
there is no spreading of $P_{{\rm S_F},0}(i,j)$ 
into off-diagonal regions with increasing system size, and 
the smoothed distribution function $\overline{P}_{{\rm S_F},0}(r)$ 
displays the same trends as before.

So far, every state investigated
can be regarded as a bound electron-hole pair with different degrees of
localization. 
They are molecular analogues of the excitons in the physics
 of semiconductors.
In order to see the differences to a 
pure ``band excitation'', the analysis is repeated for  
$|\Phi_{\rm H \to L}\rangle$, see~(\ref{HOMOLUMO}). 
As expected, and confirmed in Figures~\ref{Fig:8a}, \ref{Fig:8b}, and
\ref{Fig:8c}, 
the motion of electron and hole in the molecule is
uncorrelated. In contrast to the excitonic cases before, 
the almost linear increase of the
mean electron-hole separation and its variance as a function
of system size in Figure~\ref{Fig:8a}
is accompanied by a broadening and flattening of the
smoothed distribution function $\overline{P}_{{\rm H\to L},0}(r)$,
see Figure~\ref{Fig:8b}. As seen from Figure~\ref{Fig:8c}
there is considerable weight in the off-diagonal region in 
the probability distribution $P_{{\rm H\to L},0}(i,j)$,
which actually looks like a half sphere. 
Hence, the excitonic
behavior of the states $\rm S_1$, $\rm S_m$, and $\rm S_F$ is genuine,
and not just a coincidence in the uncorrelated motion of 
an electron and a hole in a restricted geometry.

\section{Conclusion}
\label{Conclusions}

The ground and singlet excited states of
various oligomers of LPPP have been described with
semiempirical methods. A qualitative agreement was achieved
with the experimental absorption spectra, especially for the 
measurements of the 
fluorescence lifetime, and the mean 
electron-hole distance. A new analysis
of the excited states has been given. As in optical
absorption experiments, the overlap matrix elements between 
excited states and the ground state with an electron-hole
pair are studied as a function of their respective positions
on the oligomers.
Excited states with high oscillator strengths are found to be
bound electron-hole pairs. Therefore, no
explanation of the abatic onset of the photocurrent 
in LPPP films can be given
at this level of a microscopic theory.

\acknowledgments

This work has been made possible 
with the kind technical support of Prof.~Frenking,
and the financial support of the 
Graduiertenkolleg ``Optoelektronik mesoskopischer Halbleiter'' 
as well as the AURC and the Marsden Fund in Wellington.

\begin{figure}
\caption{The LPPP oligomers; the substituents used by
various experimental groups are summarized in Table~\protect\ref{Tab:0}. 
In this work, all substituents are replaced by hydrogen.}
\label{Fig:1}
\end{figure}


\begin{figure}[h]
\caption{Optimized geometric structure of the n = 0.5 oligomere of LPPP on MP2/6-31G* level of theory.}
\label{Fig:1.5}
\end{figure}

\begin{figure}[h]
\caption{Calculated absorption spectra as a function of the oligomer length.}
\label{Fig:2}
\end{figure}


\begin{figure}[h]
\caption{Calculated energetic position of the optical 
transitions depending on 
the system size, compare Table~\protect\ref{Tab:3}. 
The experimental values 
for the $\rm S_1\leftarrow S_0$ transition are taken 
from~\protect\cite{pauck}. }
\label{Fig:3a}
\end{figure}


\begin{figure}[h]
\caption{Calculated fluorescence lifetimes as a function of the system size,
compare Table~\protect\ref{Tab:3}.}
\label{Fig:3b}
\end{figure}


\begin{figure}[h]
\caption{For plotting $P_{s,0}(i,j)$ 
along a quasi one-dimensional chain, 
only those carbon atoms are into account which are 
marked with bold lines.}
\label{Fig:4a}
\end{figure}

\begin{figure}[h]
\caption{$P_{{\rm S_1},0}(i,j)\equiv P_{ij}$ 
for the $\rm S_1$ state of the oligomer with 
two phenylene rings ($n=0.5$). 
The state is determined by finite-size 
effects of the molecule, i.e., there is no bound electron-hole pair.}
\label{Fig:4b}
\end{figure}


\begin{figure}[h]
\caption{$P_{{\rm S_1},0}(i,j)\equiv P_{ij}$ 
for the $\rm S_1$ state of the oligomer with 
five phenylene rings ($n=2$). A bound Frenkel exciton is discernible.}
\label{Fig:4c}
\end{figure}


\begin{figure}[h]
\caption{Expectation value 
$\langle E\rangle_{{\rm S_1},0}$ and standard deviation $\sigma_{{\rm S_1},0}$ 
for the state $\rm S_1$ in units of~{\AA} as a function of 
the oligomer length.}
\label{Fig:5a}
\end{figure}

\begin{figure}[h]
\caption{Smoothed distribution function
$\overline{P}_{{\rm S_1},0}(r)\equiv \overline{P}_{ij}$ as a function of the separation $r$ 
for the state $\rm S_1$ for various oligomer lengths.}
\label{Fig:5b}
\end{figure} 


\begin{figure}[h]
\caption{Probability distribution
$P_{{\rm S_1},0}(i,j)\equiv P_{ij}$ along a quasi one-dimensional 
path for the oligomers with eleven ($n=5$) phenylene rings
for the state $\rm S_1$.}
\label{Fig:5c}
\end{figure}


\begin{figure}[h]
\caption{Expectation value 
$\langle E\rangle_{{\rm S_m},0}$ and standard deviation $\sigma_{{\rm S_m},0}$ 
for the state $\rm S_m$ in units of~{\AA} as a function of 
the oligomer length.}
\label{Fig:6a}
\end{figure}

\begin{figure}[h]
\caption{Smoothed distribution function
$\overline{P}_{{\rm S_m},0}(r)\equiv \overline{P}_{ij}$ as a function of the separation $r$ 
for the state $\rm S_m$ for various oligomer lengths.}
\label{Fig:6b}
\end{figure}


\begin{figure}[h]
\caption{Probability distribution
$P_{{\rm S_m},0}(i,j)\equiv P_{ij}$ 
along a quasi one-dimensional 
path for the oligomers with eleven ($n=5$) phenylene rings
for the state $\rm S_m$.}
\label{Fig:6c}
\end{figure}


\begin{figure}[h]
\caption{Expectation value 
$\langle E\rangle_{{\rm S_F},0}$ and standard deviation $\sigma_{{\rm S_F},0}$ 
for the state $\rm S_F$ in units of~{\AA} as a function of 
the oligomer length.}
\label{Fig:7a}
\end{figure}

\begin{figure}[h]
\caption{Smoothed distribution function
$\overline{P}_{{\rm S_F},0}(r)\equiv \overline{P}_{ij}$ as a function of the separation $r$ 
for the state $\rm S_F$ for various oligomer lengths.}
\label{Fig:7b}
\end{figure}


\begin{figure}[h]
\caption{Probability distribution
$P_{{\rm S_F},0}(i,j)\equiv P_{ij}$ along a quasi one-dimensional 
path for the oligomers with eleven ($n=5$) phenylene rings
for the state $\rm S_F$.}
\label{Fig:7c}
\end{figure}


\begin{figure}[h]
\caption{Expectation value 
$\langle E\rangle_{{\rm H \to L},0}$ and standard deviation 
$\sigma_{{\rm H \to L},0}$ 
for the state $|\Phi_{\rm H \to L}\rangle$ 
in units of~{\AA} as a function of 
the oligomer length.}
\label{Fig:8a}
\end{figure}

\begin{figure}[h]
\caption{Smoothed distribution function
$\overline{P}_{{\rm H\to L},0}(r) \equiv \overline{P}_{ij}$ as a function of the separation $r$ 
for the state $|\Phi_{\rm H \to L}\rangle$ 
for various oligomer lengths.}
\label{Fig:8b}
\end{figure}


\begin{figure}[h]
\caption{Probability distribution
$P_{{\rm H\to L},0}(i,j)\equiv P_{ij}$ along a quasi one-dimensional 
path for the oligomers with with (a) two ($n=0.5$) and
(b) eleven ($n=5$) phenylene rings
for the state $|\Phi_{\rm H \to L}\rangle$.}

\label{Fig:8c}
\end{figure}

\begin{table}
\caption{Aliphatic substituents as defined in figure~\protect\ref{Fig:1}.}
\begin{tabular}{c|ccc}
 & \cite{harrison}, \cite{siggi} & \cite{grimme}, \cite{pauck} & this work\\ \hline
$\rm R_1$ & n-Hexyl& n-Hexyl & H \\
$\rm R_2$ & p(n-Decyl)-phenyl& p(t-Butyl)-phenyl & H \\
$\rm R_3$ & Methyl & H & H\\ \hline
$\rm R'_1$ & \multicolumn{3}{c}{H if terminus, $\rm R_1$ otherwise}
\end{tabular}
\label{Tab:0}
\end{table}

\begin{table}
\caption{Comparison between MP2, AM1, and PM3.}
\begin{tabular}{c|cccc}
Method & E / a.u. \tablenote{ E = electronic energy (MP2), heat of formation (AM1, PM3)} &
ZPE/(kcal $\cdot$ mol$^{-1}$) \tablenote{ ZPE= zero point energy}
&
point group &
$\sum|\Delta r_{\rm C-C}|$ \tablenote{ $\sum|\Delta r_{\rm C-C}|$ 
= Sum of the differences of the C--C bond 
lengths with respect to the optimized structure at MP2 level}\\ \hline
MP2/6-31G* & -499.8413498& 117.80 & $C_{2v}$ & - \\
AM1 & 0.0862897 & 121.55  & $C_{2v}$  & 0.000 \\
PM3 & 0.0778001 & 117.65 & $C_{2v}$ & 0.029 
\end{tabular}
\label{Tab:1}
\end{table}

\begin{table}
\caption{Ground state energies of the LPPP oligomers at AM1 level.}
\begin{tabular}{l|cccc}
N \tablenote{ N = number of phenylene rings} 
& E/a.u. \tablenote{ E = heat of formation} & 
ZPE/(kcal $\cdot$ mol$^{-1}$) \tablenote{ ZPE= zero point energy}
& E + ZPE /a.u. & point group\\
\hline
2 & 0.0862897  & 121.55  & 0.279986  & $C_{2v}$\\

3 &  0.1373472  &  178.64  &  0.422035  & $C_{2h}$ \\

4 &  0.1883913  &  235.71  &  0.564020  & $C_{2v}$ \\

5 &  0.2394355  &  292.76  &  0.705985  & $C_{2h}$ \\

7 & 0.3415265  &  406.85  &  0.989898  & $C_{2h}$ \\

9 & 0.4436183  & 520.95  &  1.273804  & $C_{2h}$ \\

11 & 0.5457102  & 635.03  &  1.557695  & $C_{2h}$ 
\end{tabular}
\label{Tab:2}
\end{table}

\begin{table}
\caption{Calculated energetic position of the transitions 
from the ground state in cm$^{-1}$ 
to the $\rm S_1$, $\rm S_m$, and $\rm S_F$ state, 
depending on the system size, and fluorescence 
lifetimes depending on the system size. 
The experimental values for the 
$\rm S_1\leftarrow S_0$ transition are taken from~\protect\cite{pauck}.}

\begin{tabular}{c|cccc|c}
$1/n$ \tablenote{ $n$ = number of monomer units} & Experiment~\cite{pauck}& 
$\rm S_1$ & $\rm S_m$ & $\rm S_F$ & $\tau/10^{-12}$s \tablenote{ $\tau$ = 
fluorescence lifetime following equation~(\protect\ref{lifetime})}
\\ \hline
0.500 & - & 34745.8 & - & 47703.1 & 1789 \\
$0.\bar{3}$ & 29500 & 31214.5 & -& 47629.4 & 1683 \\
0.250 & - & 29289.2 & - & 47145.2 & 1115 \\
0.200 & 25700 & 28055.1 & 3861.4 & 46995.0 & 920 \\
0.143 & 23800 & 26841.0 & 34345.4 & 46851.2 & 693 \\
$0.\bar{1}$ & -& 26187.2 & 31613.2 & 46758.9 & 560 \\
$0.\bar{09}$ & -& 25795.0 & 30272.1 & 46686.4 & 470 \\
$0.08\bar{3}$ & 22100 & - & - & - & - 
\end{tabular}
\label{Tab:3}
\end{table}

\begin{table}
\caption{Values of the expectation value $\langle E\rangle_{s,0}$ 
and the standard orientation $\sigma_{s,0}$ 
for the optically detectable states, 
depending on the size of the oligomers.}
\begin{tabular}{lc|ccccccc}
 & & \multicolumn{7}{c}{N \tablenote{ N = number of phenylene rings}}\\
{ state} & 
{values \tablenote{ [$\langle E\rangle$ ]=[$\sigma$]= 1 
{\AA}} }& 2 & 3 & 4 & 5 & 7 & 9 & 11 \\ \hline
$|\Phi_{\rm H\to L}\rangle$ & $\langle E\rangle_{{\rm H\to L},0}$  
& 2.93 & 3.86 & 4.76 & 5.64 & 9.01 & 9.07 & 10.79 \\
     &$\sigma_{{\rm H\to L},0}$ & 1.80 & 2.49 & 3.16 & 3.82 & 5.77 & 6.38 & 7.65 \\ \hline
$\rm S_1$ &$\langle E\rangle_{{\rm S_1},0}$ & 2.34 & 2.96 & 3.37 & 3.58 & 3.79 & 3.87 & 3.90 \\
      &$\sigma_{{\rm S_1},0}$ & 1.13& 1.91 & 2.35 & 2.59 & 2.82 & 2.91 & 2.96 \\ \hline
$\rm S_m$ & $\langle E\rangle_{{\rm S_m},0}$  & -& -& -& 3.06 & 3.05 & 3.20 & 3.52 \\
       &$\sigma_{{\rm S_m},0}$ & -& -& -& 2.18 & 2.19 & 2.21 & 2.59 \\ \hline
$\rm S_F$ & $\langle E\rangle_{{\rm S_F},0}$  & 1.99 & 2.12 & 2.74 & 2.64 & 3.90 & 3.59 & 3.80 \\
       &$\sigma_{\rm S_F,0}$ & 1.33 & 1.49 & 2.09 & 2.03 & 3.56 & 3.11 & 3.39
\end{tabular}
\label{Tab:4}
\end{table}

\newpage

\setcounter{figure}{0}

\begin{figure}
\caption{J\"org Rissler \it Phys. Rev. B }
\epsfig{file = 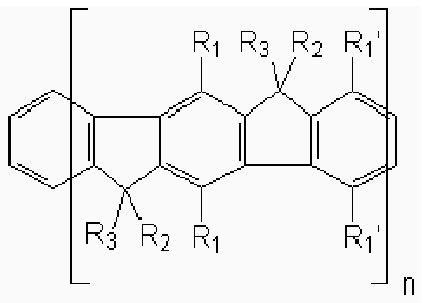, width = 4 cm}

\end{figure}


\begin{figure}
\caption{J\"org Rissler \it Phys. Rev. B }
\epsfig{file = 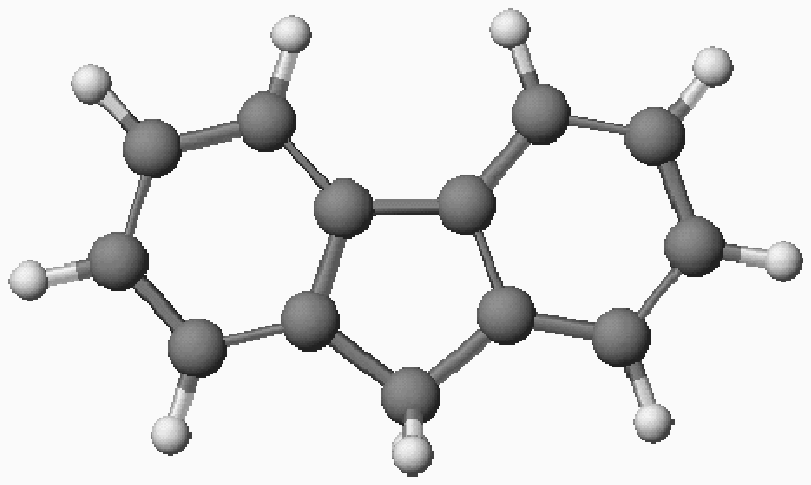, width = 6 cm}
\end{figure}


\begin{figure}
\caption{J\"org Rissler \it Phys. Rev. B }
\epsfig{file = 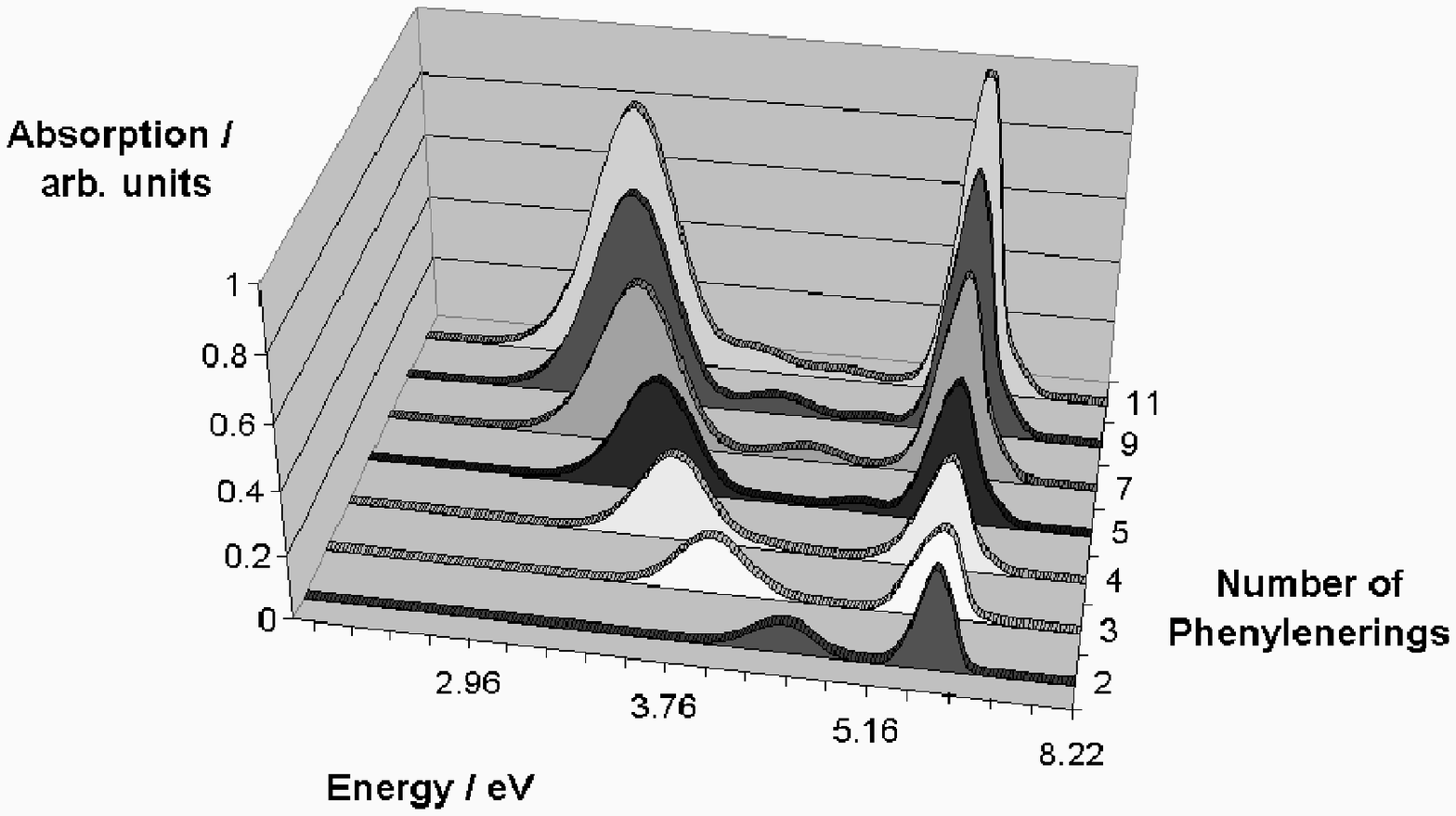, width = 8.5 cm}
\end{figure}

\newpage

\begin{figure}
\caption{J\"org Rissler \it Phys. Rev. B }
\epsfig{file = 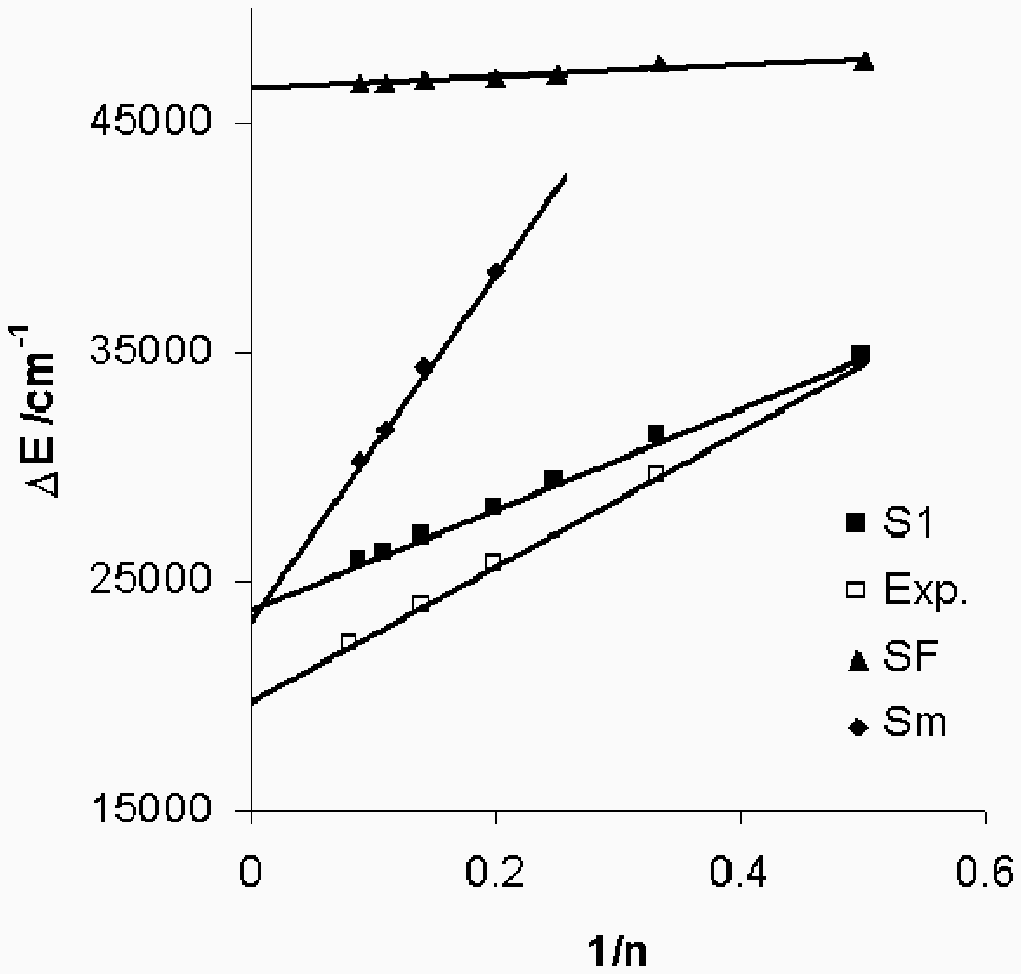, width = 8.5 cm}
\end{figure}


\begin{figure}
\caption{J\"org Rissler \it Phys. Rev. B }
\epsfig{file = 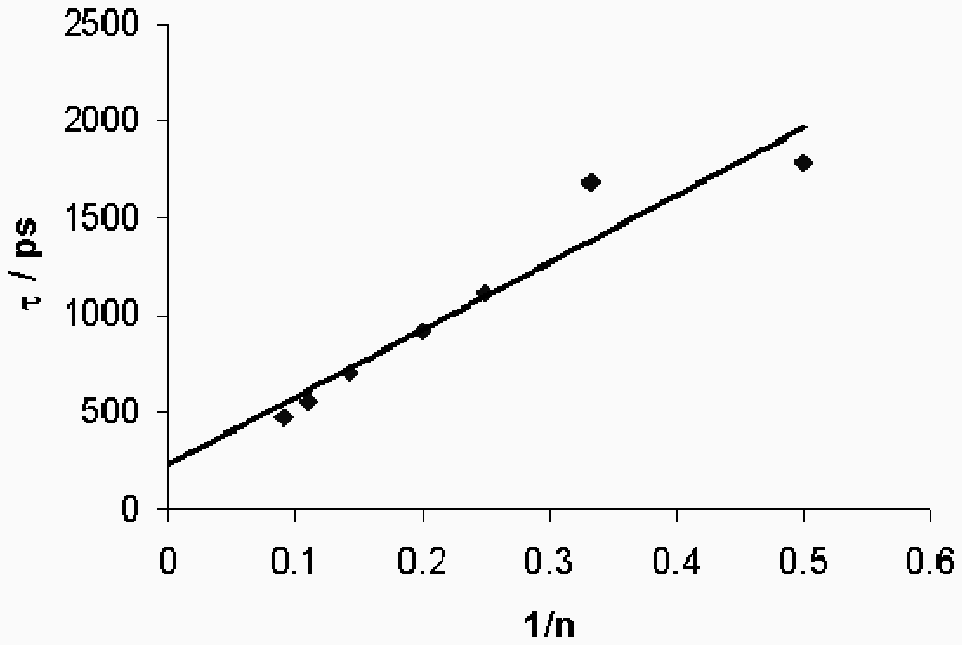, width = 8.5 cm}
\end{figure}


\begin{figure}
\caption{J\"org Rissler \it Phys. Rev. B }
\epsfig{file = 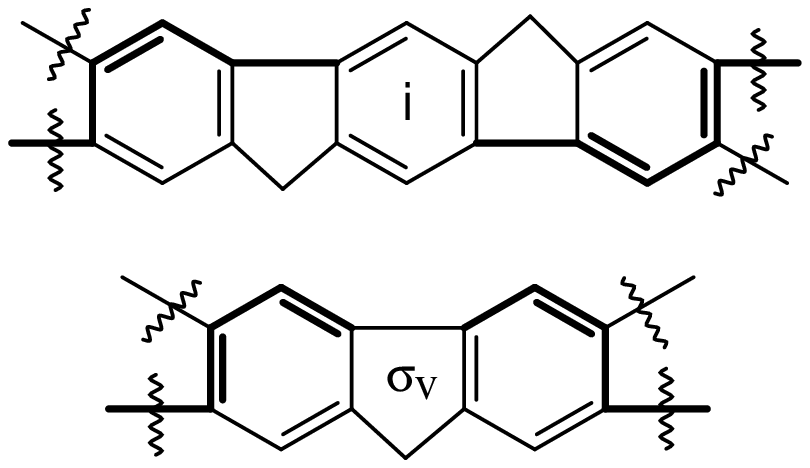, height = 5cm}
\end{figure}

\newpage

\begin{figure}
\caption{J\"org Rissler \it Phys. Rev. B }
\epsfig{file = 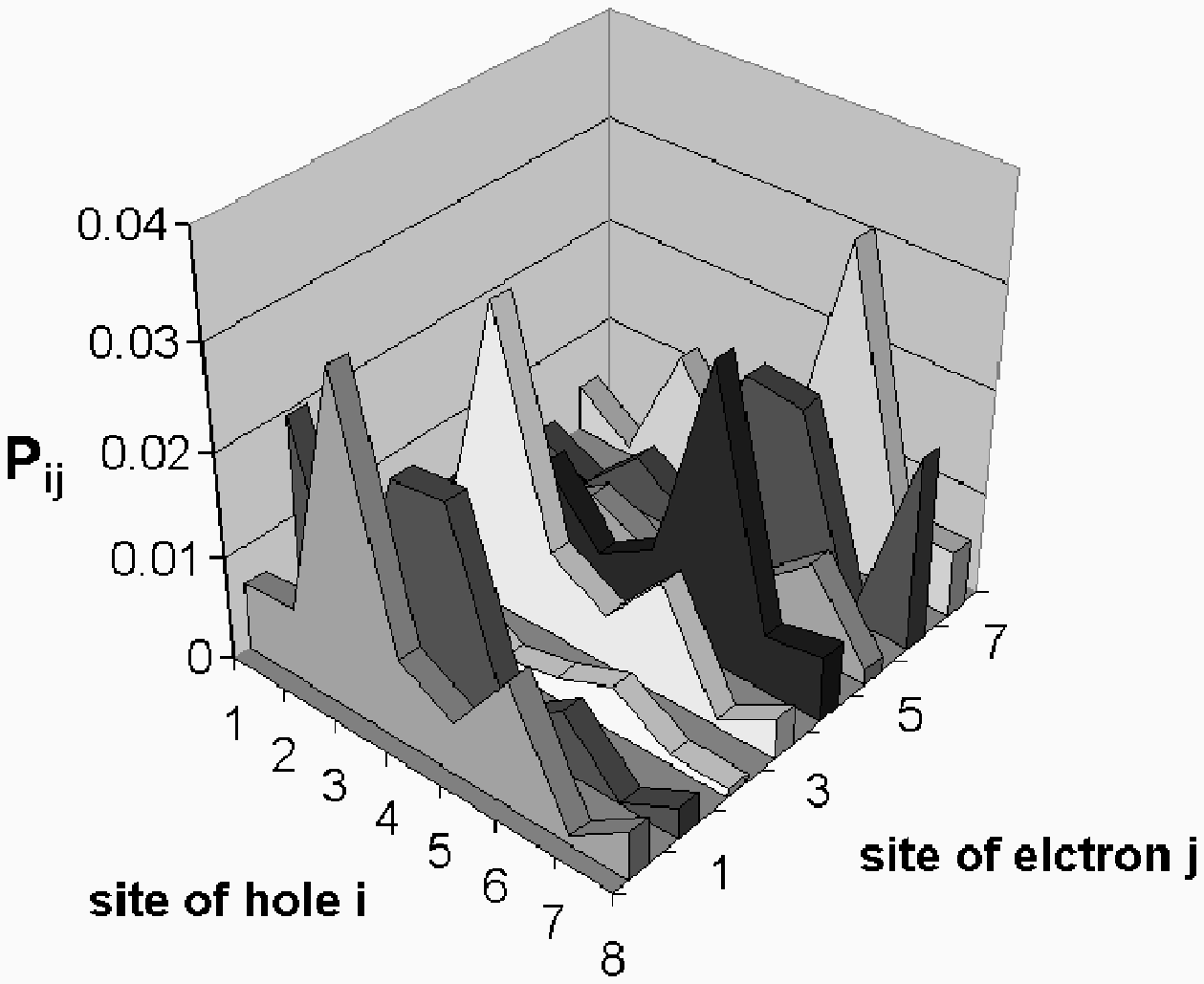, width = 8.5cm}
\end{figure}


\begin{figure}
\caption{J\"org Rissler \it Phys. Rev. B }
\epsfig{file = 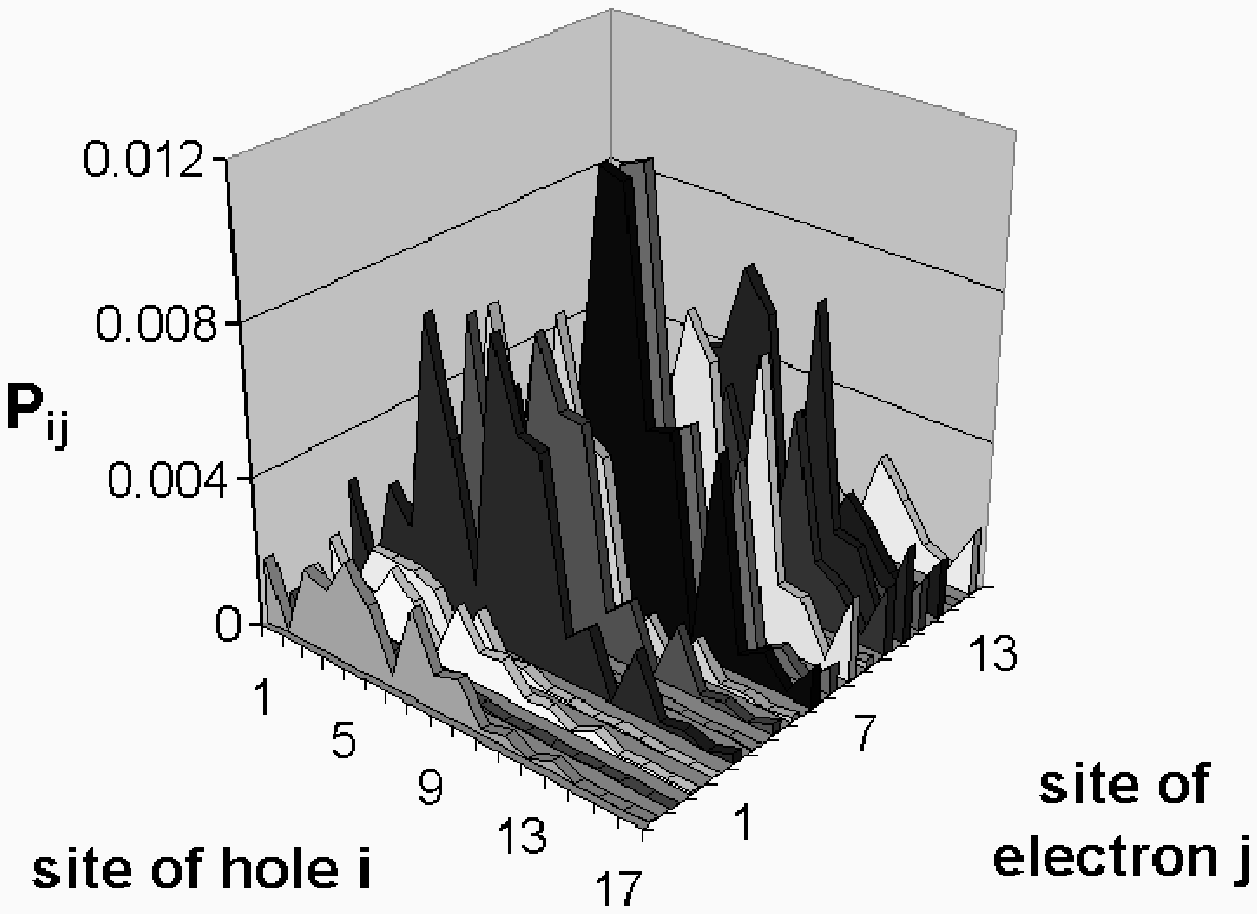, width = 8.5cm}
\end{figure}


\begin{figure}
\caption{J\"org Rissler \it Phys. Rev. B }
\epsfig{file = 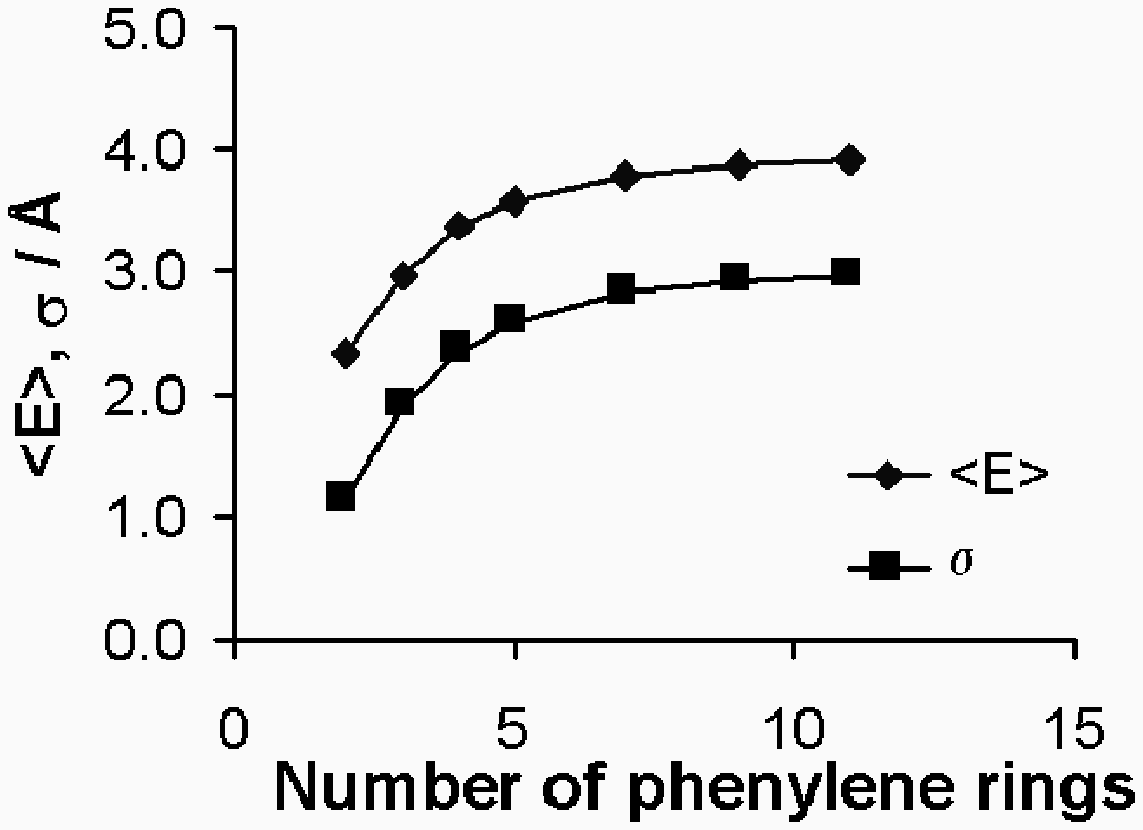, height = 5cm, width = 8.5cm}
\end{figure}

\newpage

\begin{figure}
\caption{J\"org Rissler \it Phys. Rev. B }
\epsfig{file = 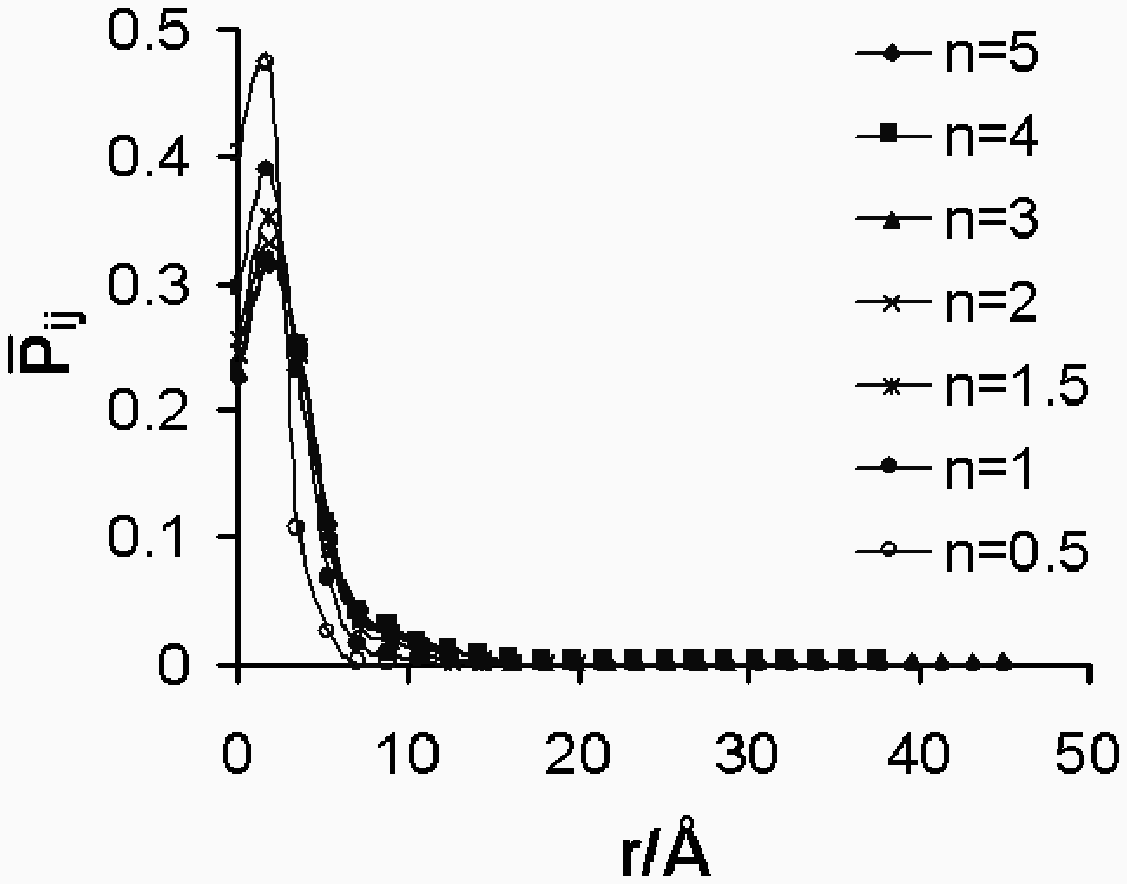, height = 5cm, width = 8.5cm}
\end{figure}


\begin{figure}
\caption{J\"org Rissler \it Phys. Rev. B }
\epsfig{file = 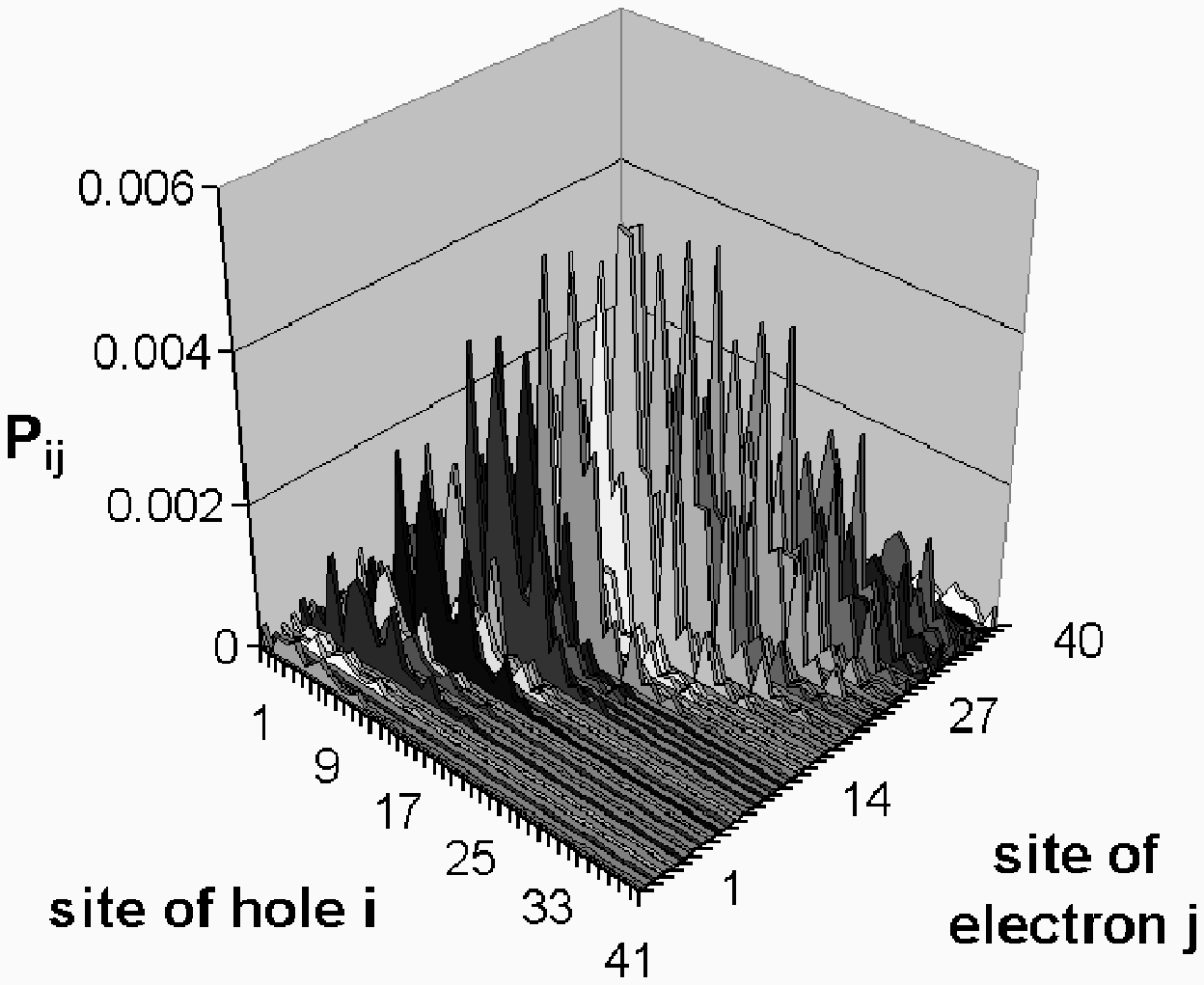, height = 5cm, width= 8.5cm}
\end{figure}


\begin{figure}
\caption{J\"org Rissler \it Phys. Rev. B }
\epsfig{file = 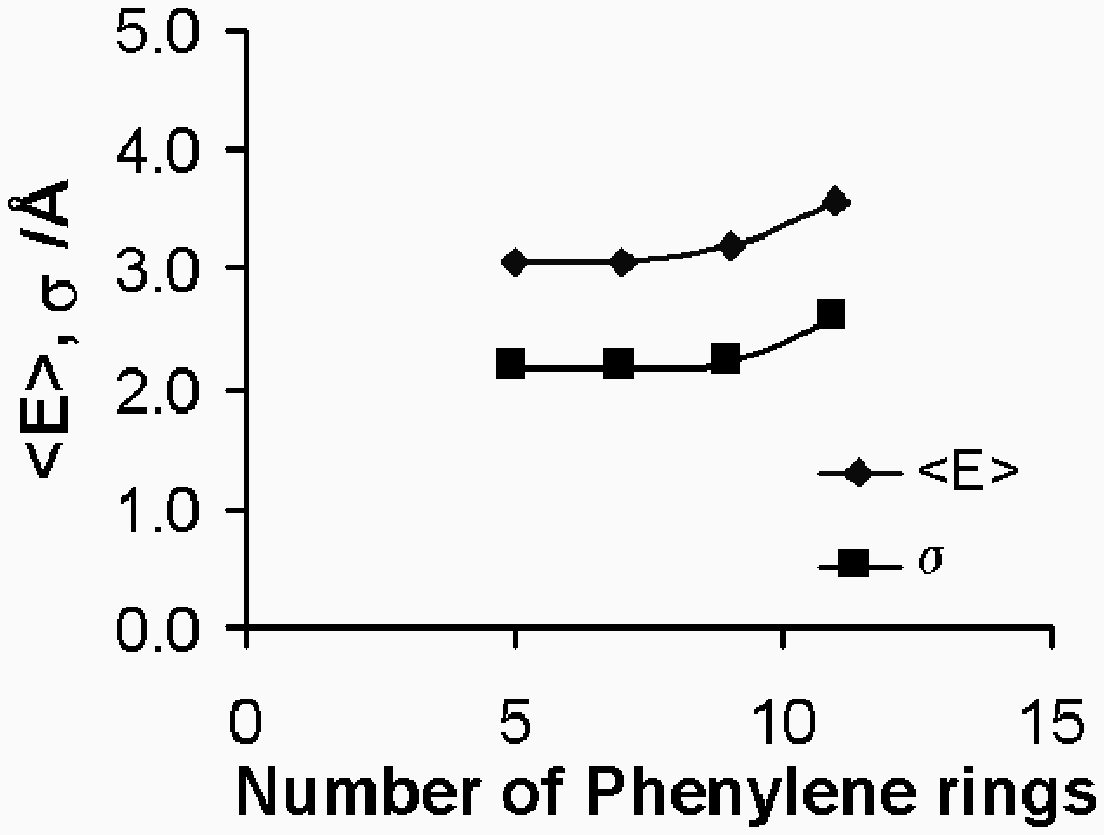, height = 5cm, width = 8.5cm}
\end{figure}

\newpage

\begin{figure}
\caption{J\"org Rissler \it Phys. Rev. B }
\epsfig{file = 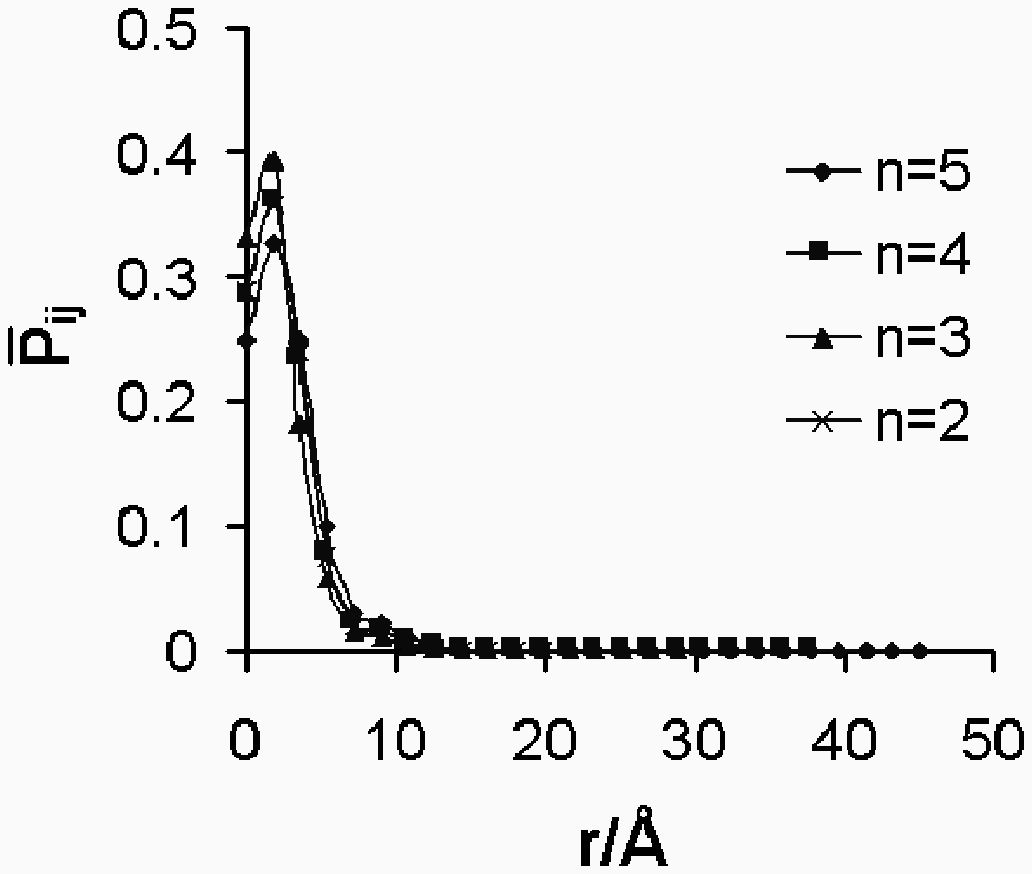, height = 5cm, width = 8.5cm}
\end{figure}


\begin{figure}
\caption{J\"org Rissler \it Phys. Rev. B }
\epsfig{file = 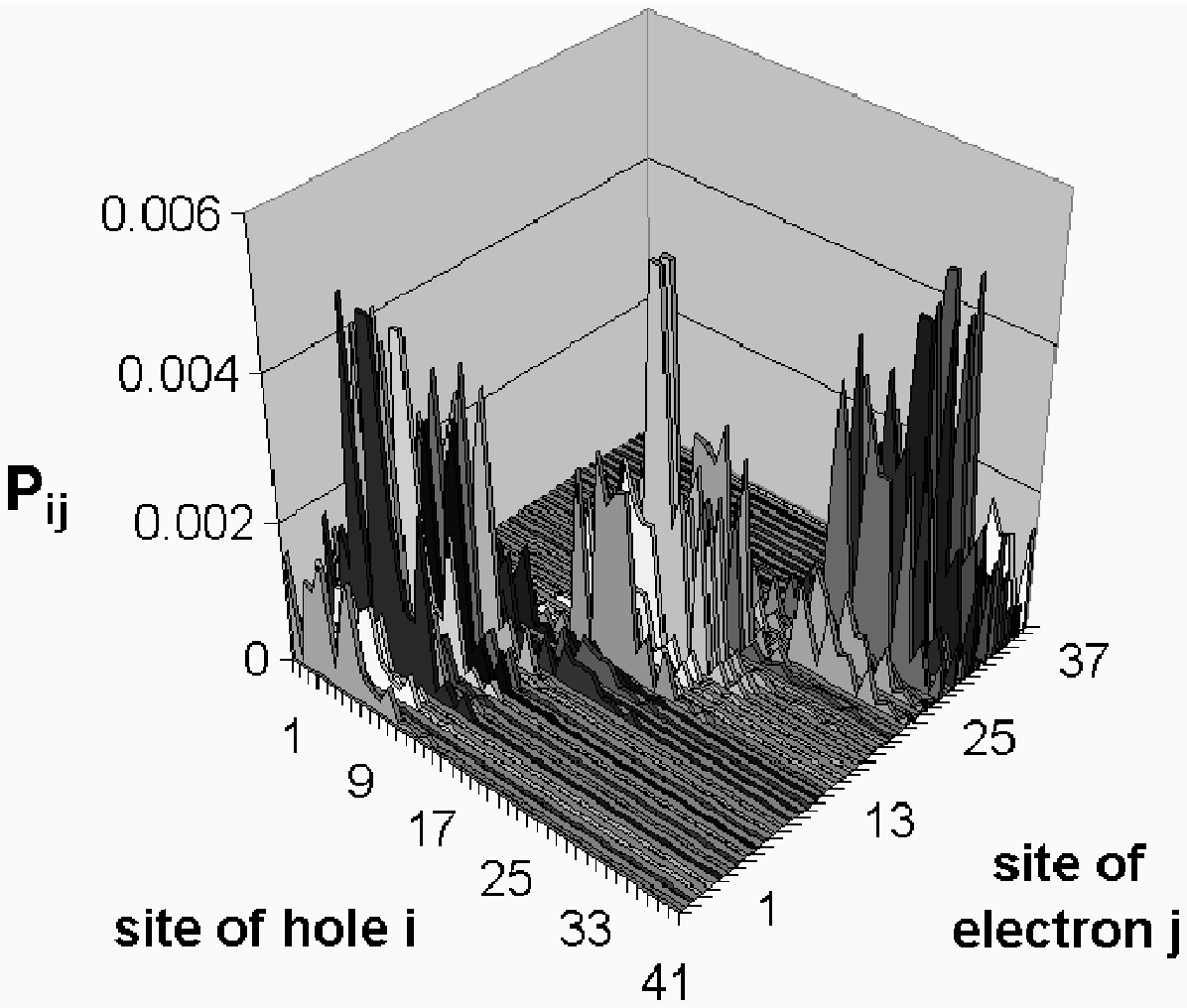, width = 8.5 cm}
\end{figure}


\begin{figure}
\caption{J\"org Rissler \it Phys. Rev. B }
\epsfig{file = 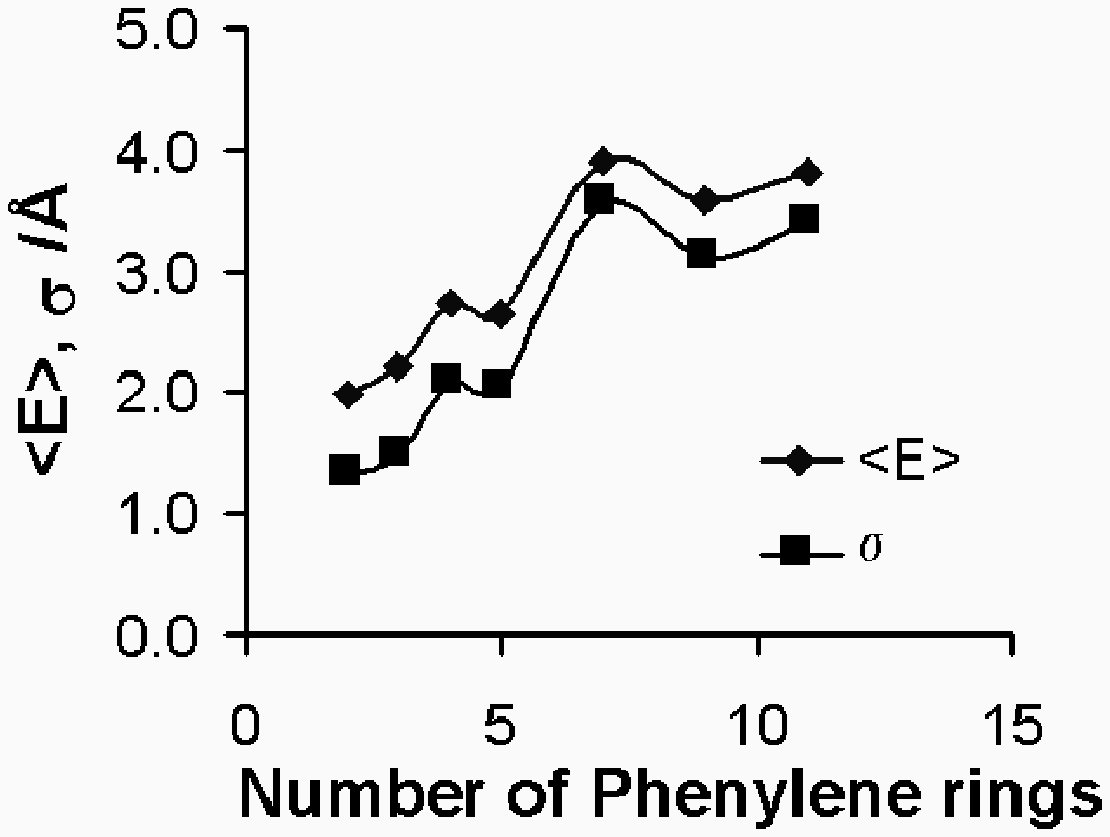, height = 5cm, width = 8.5cm}
\end{figure}

\newpage

\begin{figure}
\caption{J\"org Rissler \it Phys. Rev. B }
\epsfig{file = 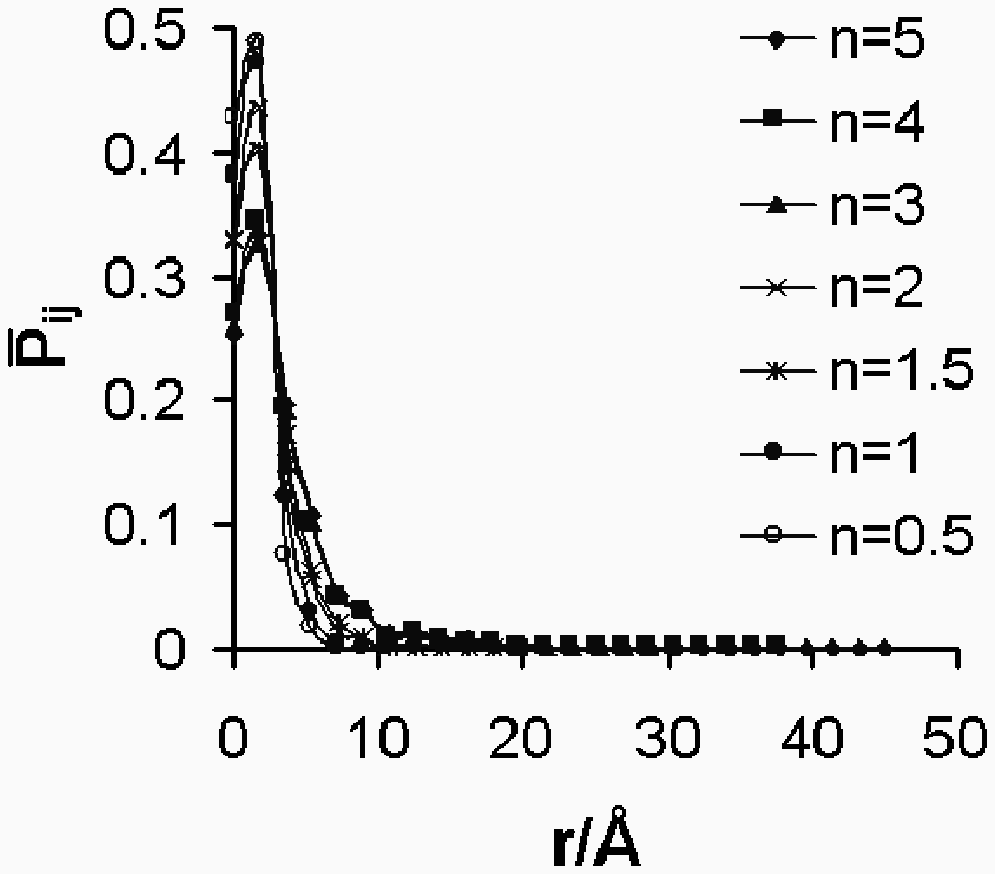, height = 5cm, width = 8.5cm}
\end{figure}


\begin{figure}
\caption{J\"org Rissler \it Phys. Rev. B }
\epsfig{file = 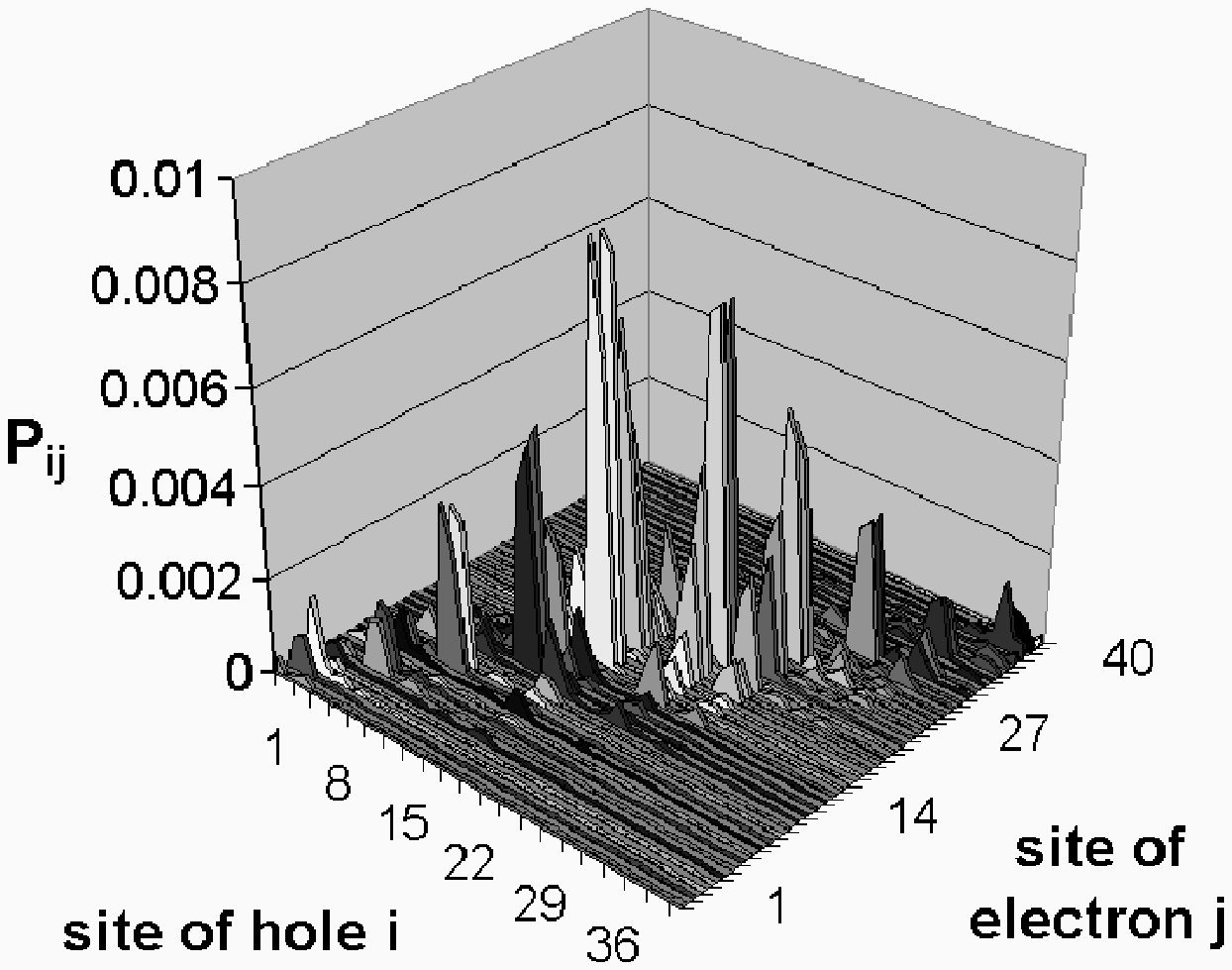, width = 8.5 cm}
\end{figure}


\begin{figure}
\caption{J\"org Rissler \it Phys. Rev. B }
\epsfig{file = 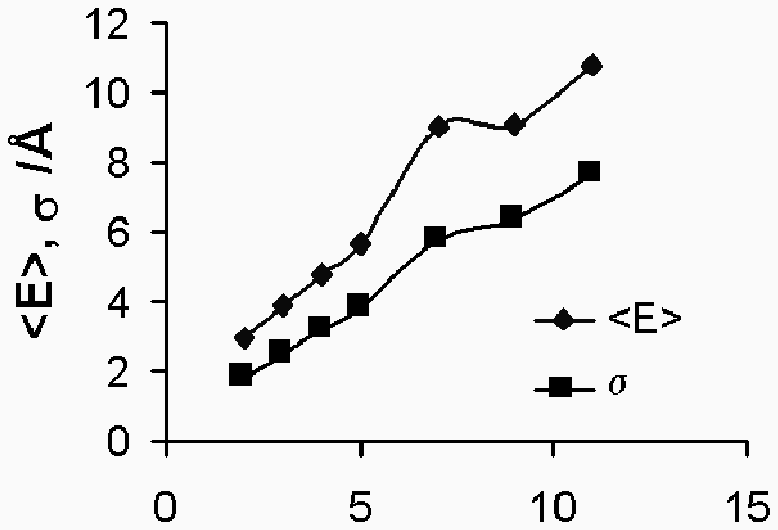, height = 5cm, width = 8.5cm}
\end{figure}

\newpage

\begin{figure}
\caption{J\"org Rissler \it Phys. Rev. B }
\epsfig{file = 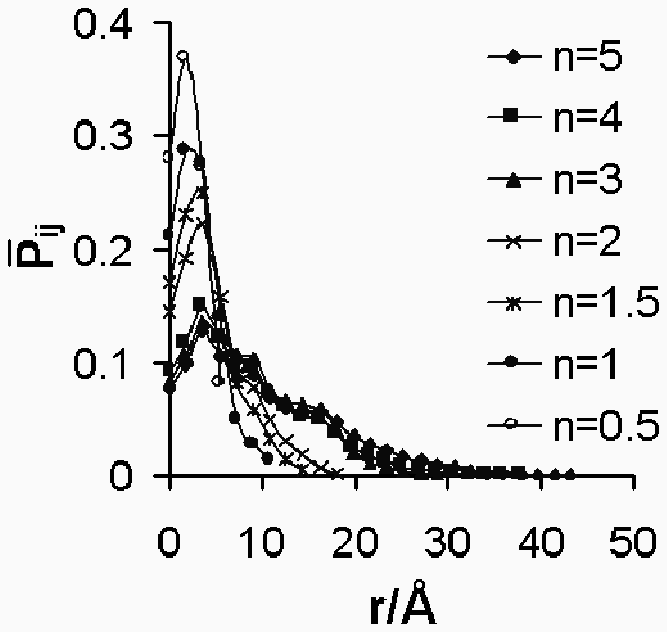, height = 5cm, width = 8.5cm}
\end{figure}


\begin{figure}
\caption{J\"org Rissler \it Phys. Rev. B}
a) \epsfig{file = 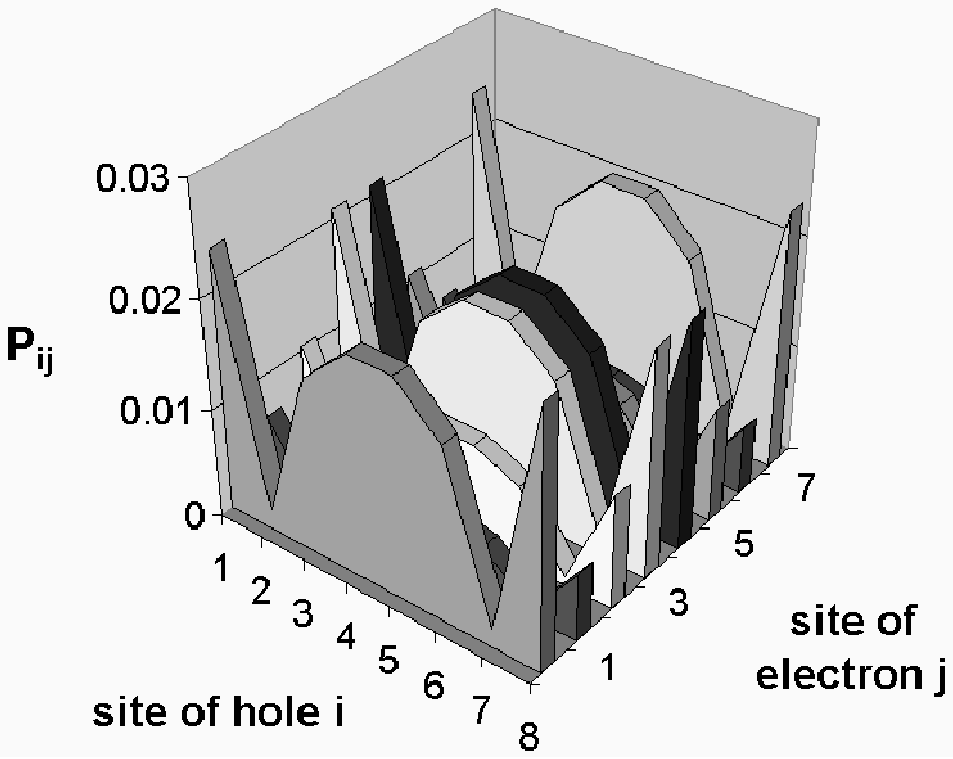, width =8.5 cm}\\
b) \epsfig{file = 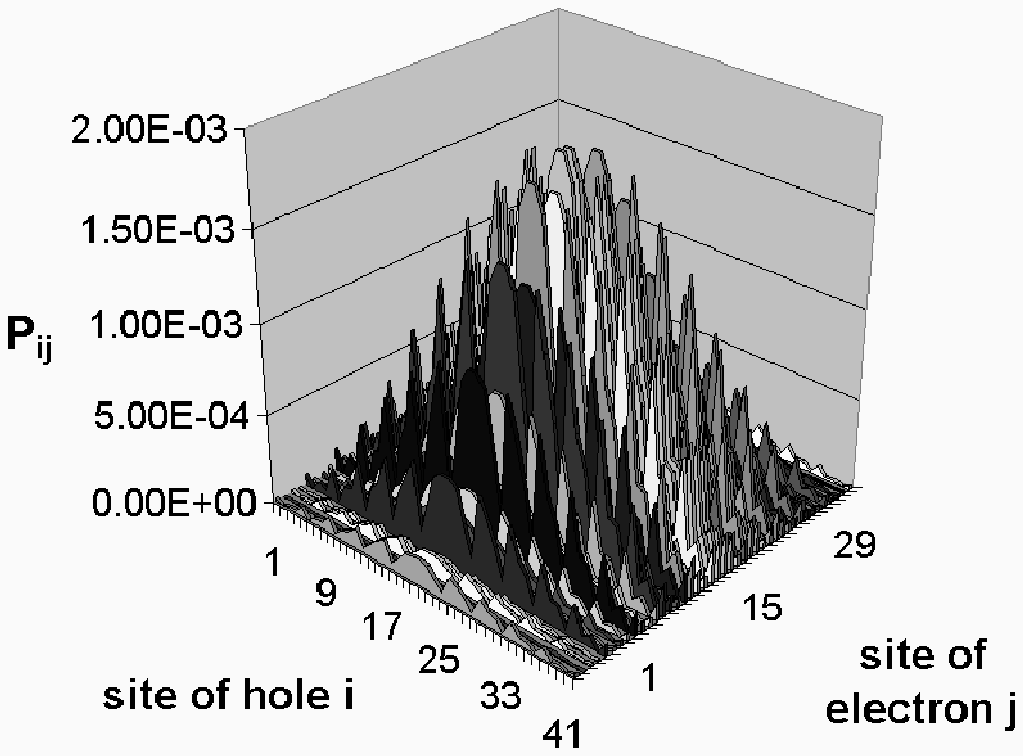,  width = 8.5 cm}
\end{figure}


\begin{references}

\bibitem{scherf} U. Scherf, A. Bohnen and K. M\"ullen, 
\textit{Macromol. Chem.}  \textbf{193}, 1127 (1992).

\bibitem{harrison} M. G. Harrison, S. M\"oller, 
G. Weiser, G. Urbasch, R. F. Mahrt and H. B\"assler, 
{\it Phys. Rev. B} \textbf{60}, 8650 (1999).

\bibitem{olig}  G. Wegner and K. M\"ullen, Electronic Materials: The Oligomer Approach, 
(Wiley/VCH, Weinheim, 1996).

\bibitem{grimme} J. Grimme, M. Kreyenschmidt, F. Uckert, K. M\"{u}llen,  
and U. Scherf, \textit{Adv. Mater.} \textbf{7}, 292 (1995).

\bibitem{pauck} T. Pauck, H. B{\"a}ssler, J. Grimme, U. Scherf and  
K. M{\"u}llen, \textit{Chem. Phys.} \textbf{210}, 219 (1996).

\bibitem{siggi} S. Barth, H. B\"{a}ssler, U. Scherf and K. M\"{u}llen,
\textit{Chem. Phys. Lett.} \textbf{288}, 147 (1998).

\bibitem{alvarado} S. F. Alvarado, P. F. Seidler, D. G. Lidzey,  
and D. D. G. Bradley, {\it Phys. Rev. Lett.} \textbf{81}, 1082 (1998).

\bibitem{bredas} A. K\"{o}hler, D. A. dos Santos, D. Beljonne,  
 Z. Shuai, J.-L. Br\'edas, A. B. Holmes,  A. Kraus,  K. M\"{u}llen, 
 and R. H. Friend, \textit{Nature} \textbf{392}, 903 (1998).

\bibitem{archipov} V. I. Archipov, E. V. Emelianova and H. B{\"a}ssler,  
{\it Phys. Rev. Lett}  {\bf 82}, 1321 (1999).

\bibitem{pople}
J. A. Pople, D. L. Beveridge and P. A. Dobosh, 
{\it J. Chem. Phys.} \textbf{47}, 2026 (1967).


\bibitem{MP2} C. M\"{o}ller and M. S. Plesset,
\textit{Phys. Rev.} \textbf{46}, 618 (1934).


\bibitem{basis} R. Ditchfield, W. J. Hehre and J. A. Pople, 
\textit{J. Chem. Phys.} \textbf{54}, 724 (1971).\\
 W. J. Hehre, R. Ditchfield and J. A. Pople, \textit{J. Chem. Phys.}  
\textbf{56}, 2257 (1972).\\
 P. C. Hariharan and J. A. Pople, \textit{Mol. Phys.}  
\textbf{27}, 209 (1974).\\
 M. S. Gordon, \textit{Chem. Phys. Lett.} \textbf{76}, 163 (1980).\\
 P. C. Hariharan and J. A. Pople, \textit{Theo. Chim. Acta}
\textbf{28}, 213 (1973).\\
 G. A. Petersson,  A. Bennett, T. G. Tensfeldt,  M. A. Al-Laham,
W. A. Shirley and J. Mantzaris, \textit{J. Chem. Phys.} 
 \textbf{89}, 2193 (1988).

\bibitem{bredas2} J. Cornil, D. Beljonne, D. A. dos Santos, 
Z. Shuai and J.-L. Br\'edas, \textit{Synth. Met.}  
\textbf{78}, 209 (1996).

\bibitem{dewar} J. S. Dewar, E. G. Zoebisch and E. F. Healy, \textit{J. Am. Chem. Soc.} \textbf{107}, 3902 (1985).

\bibitem{stewart} J. J. P. Stewart, \textit{J. Comp. Chem.}  \textbf{10}, 221 (1989).

\bibitem{gauss} Gaussian 94, Revision D4;
 M.J. Frisch, G.W. Trucks, H.B. Schlegel, G.E. Scuseria,
 M.A. Robb, J.R. Cheeseman, V.G. Zakrzewski, J.A. Montgomery, Jr.,
 R.E. Stratmann, J.C. Burant, S. Dapprich, J.M. Millam,
 A.D. Daniels, K.N. Kudin, M.C. Strain, O. Farkas, J. Tomasi,
 V. Barone, M. Cossi, R. Cammi, B. Mennucci, C. Pomelli, C. Adamo,
 S. Clifford, J. Ochterski, G.A. Petersson, P.Y. Ayala, Q. Cui,
 K. Morokuma, D.K. Malick, A.D. Rabuck, K. Raghavachari,
 J.B. Foresman, J. Cioslowski, J.V. Ortiz, A.G. Baboul,
 B.B. Stefanov, G. Liu, A. Liashenko, P. Piskorz, I. Komaromi,
 R. Gomperts, R.L. Martin, D.J. Fox, T. Keith, M.A. Al-Laham,
 C.Y. Peng, A. Nanayakkara, C. Gonzalez, M. Challacombe,
 P.M.W. Gill, B. Johnson, W. Chen, M.W. Wong, J.L. Andres,
 C. Gonzalez, M. Head-Gordon, E.S. Replogle and J.A. Pople,
 Gaussian, Inc., Pittsburgh PA, 1998.

\bibitem{zerner} J. E. Ridley and M. C. Zerner,
\textit{Theo. Chim. Acta} \textbf{32}, 111 (1973).\\
M. C. Zerner, {\it Semiempirical Molecular Orbital Methods}
in \textit{Reviews of Computational Chemistry}, edited by K. B. Lipkowitz and
D. B. Boyd (VCH Publishing, New York, 1991), vol.~2, pp.~313-365.

\bibitem{cache} CAChe 3.1, Oxford Molecular Ltd., 1997.

\bibitem{iura} Yu. N. Romanovskii (unpublished).

\bibitem{mahan} See, e.g., G. D. Mahan, {\it Many-Particle Physics}, 
2nd edition (Plenum Press, New York, 1990).

\bibitem{gw} M. Rohlfing and S. G. Louie, \textit{Phys. Rev. Lett.} 
 \textbf{82}, 1959 (1999).\\
J.-W. van der Horst, P. A. Bobbert, P. H. L. de Jong, M. A. J. Michels, 
G. Brocks and P. J. Kelly, \textit{Phys. Rev. B}  \textbf{61}, 
15817 (2000).

\bibitem{mazum} D. Guo, S. Mazumdar, S. N. Dixit, F.  Kajzar, F. Jarka,
Y. Kawabe and N. Peyghambarian, \textit{Phys. Rev. B} 
\textbf{48}, 1433 (1993).

\bibitem{eric} E. Jeckelmann and R. J. Bursill, (private communication).

\bibitem{knupfer} M. Knupfer, J. Fink, E. Zojer, G. Leising, 
and D. Fichou, \textit{Chem. Phys. Lett.} \textbf{318}, 585 (2000).

\end{references}
\end{document}